# The Origin and Evolution of Saturn: A Post-Cassini Perspective

SUSHIL K. ATREYA, AURÉLIEN CRIDA, TRISTAN GUILLOT, CHENG LI, JONATHAN I. LUNINE, NIKKU MADHUSUDHAN, OLIVIER MOUSIS AND MICHAEL H. WONG

## Abstract

The Saturn system has been investigated extensively since the early 1970s, with the bulk of these data generated by the Cassini-Huygens Mission between 2004 and 2017. A major thrust of those investigations has been to understand how Saturn formed and evolved and to place Saturn in the context of other gas giants and planetary systems in general. Two models have been proposed for the formation of the giant planets – the core accretion model and the disk instability model. Saturn's heavy element enrichment, core size, internal structure, etc. compared to Jupiter strongly favors the core accretion model as for Jupiter. Two features of the core accretion model that are distinct from the disk instability model are the growth of a core with a mass several times that of the Earth, followed by runaway collapse of gas onto the protoplanet once a mass threshold is reached. Current thinking is that the heavy element core grows slowly over millions of years through accretion of cm-m sized pebbles (upon sticking of grains of dust and ice particles), larger planetesimals and moon sized embryos in the gaseous disk. Gas accretion begins when the core grows to sufficient mass. The abundance pattern of heavy elements (>$^4$He) is thus a key constraint on formation models. Ratios of C/H, N/H, S/H and P/H at Saturn are presently known to varying degree of uncertainty. The He/H ratio in the atmosphere is crucial for understanding heat balance, interior processes, and planetary evolution, but present values range from low to high, allowing for a wide range of possibilities. While the very low values are favored to explain Saturn's excess luminosity, high values might indicate presence of layered convection in the interior, resulting in slow cooling. Cassini's Grand Finale orbits allowed an unprecedented opportunity to characterize Saturn's rings in exquisite detail, which together with the planet's moons, provide additional insights into formation of Saturn. While the solar system is the only analog for the extrasolar systems, detection of the alkali metals and water in giant exoplanets including exo-Saturns is proving useful for understanding the formation and evolution of Saturn, where such data are presently lacking. This chapter presents a detailed discussion of all above issues. Related material on Saturn's interior is discussed in detail in Chapter 4.

## 3.1 Introduction

Beyond its spectacular ring system, Saturn was the lesser known giant planet compared to Jupiter, both because of its twice farther distance from Earth and the more subdued appearance of its atmosphere. And yet, its very existence posed the question of why our solar system hosts two gas giants and two much smaller "icy giant" planets—Uranus and Neptune—beyond the asteroid belt. Did Jupiter and Saturn form independently and without interaction, or were they the product of complex interactions between the two? What does the existence of Uranus and Neptune tell us of this process? Exploration of the Saturn system began with the flybys of Pioneer 11 in 1979 followed by Voyager 1 and 2 in 1980 and 1981, respectively, but it was the thirteen-year tour of the Cassini-Huygens Mission from 2004-2017 that revolutionized our understanding of the planet and its environs. This chapter presents a current assessment of the understanding of the formation and evolution of Saturn gleaned from these missions, constrained by atmospheric composition and structure, internal heat flow, moons and the rings, and the latest information on the interior from Cassini (see Chapter 4 for a detailed discussion on the interior). The most comprehensive results have come from Cassini because of its extraordinarily rich scientific payload and the long duration of the mission, which allowed studying seasonal changes remotely over half of a Saturn year—from northern winter to northern summer solstice—together with direct in situ measurements down to altitudes of approximately 1700 km (<0.1 nanobar), several scale heights above Saturn's homopause (~0.01-1 µbar), in the 22 orbits of the Grand Finale phase. Figure 3.1 gives a broad overview of Saturn's atmosphere and structure, illustrating the regions explored, types of data retrieved and the unexplored regions. We discuss current models of the formation of Saturn in Sec. 3.2, followed by atmospheric composition constraints (Sec. 3.3), volatile delivery (Sec. 3.4), and, finally, insights from the moons and the rings (Sec. 3.5) and the exoplanets (Sec. 3.6).

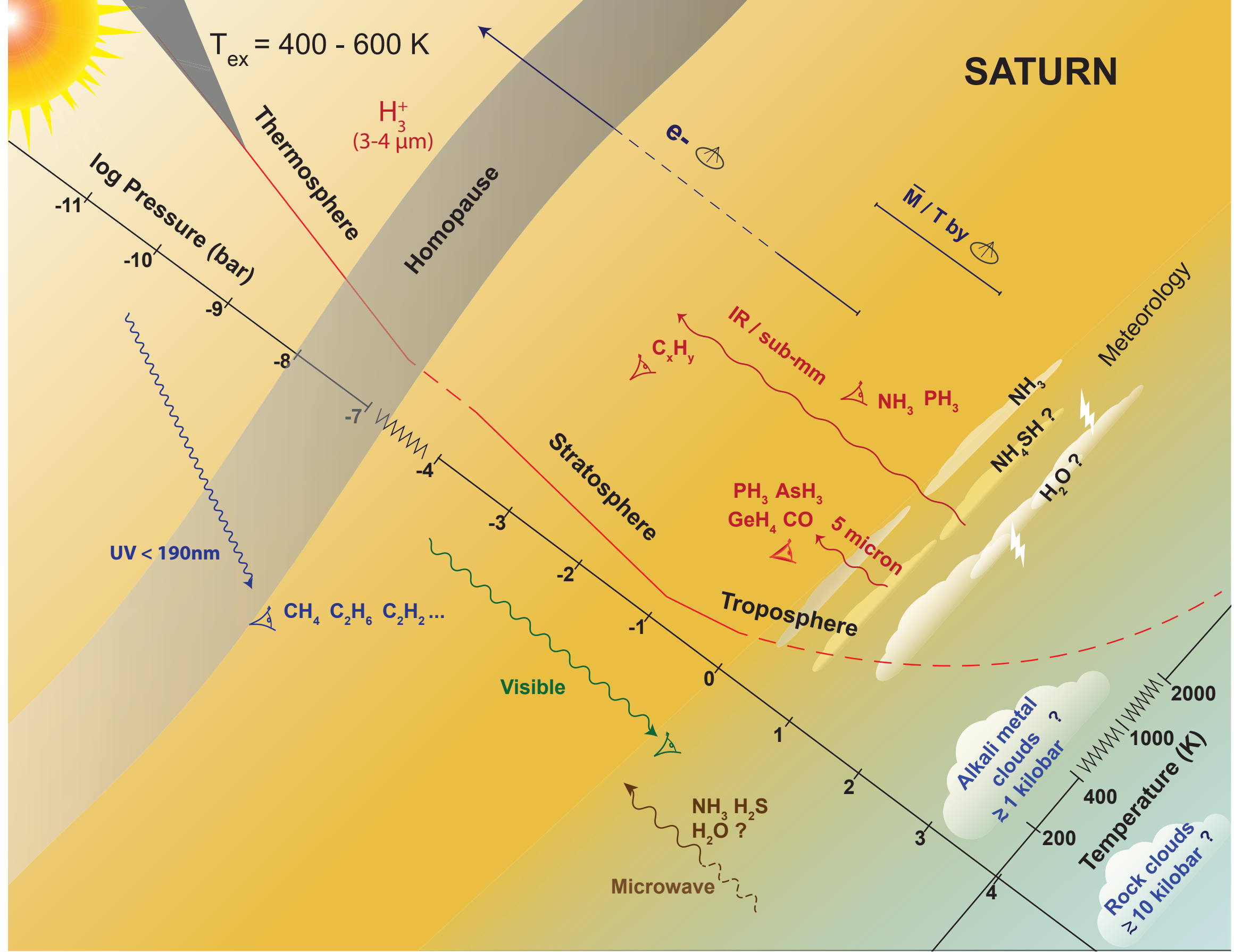

Figure 3.1 A bird's eye view of Saturn's atmospheric composition and structure, showing the regions explored, techniques, type of information retrieved, missing information (broken lines, and ?) and the cloud layers identified or predicted in the shallow and the deep troposphere. Information shown above 1 bar level is largely guided by Cassini's 13-year tour of the Saturn system. The homopause varies between ~0.01-1 µbar pressure levels across the planet. $T_{ex}$ represents the range of the exospheric temperature. e- and $\overline{M}$/T represent the electron concentration in the ionosphere and the ratio of the atmospheric mean molecular mass to temperature in the neutral atmosphere from approximately 1-1000 mb level, respectively, using the radio occultation technique. An independent knowledge of $\overline{M}$ is required to derive the absolute value of temperatures, which has been possible to do for Jupiter (Gupta et al. 2022) but not Saturn because of the lack precise measurements of Saturn's bulk composition in the absence of an entry probe, unlike Jupiter (see Sec. 3). Infrared observations allow the determination of temperatures in the stratosphere. $H_3$+ is detected in the lower ionosphere using infrared spectroscopy. 5-micron observations allow the measurement of disequilibrium species (Sec. 3.3.3). Only topmost clouds made up of $NH_3$ have been identified so far, other cloud layers in the upper troposphere are based on thermochemical model predictions (Atreya et al., 1999), as are the deeper alkali-metal and rock clouds at kilobar pressure levels (see Fig. 9 in Sec. 3.6).

## 3.2 Formation of Saturn

Three features visible by direct observation of Jupiter and Saturn suggest a difference between the formation of the two giant planets: the architectures of the satellite systems, the presence of a massive ring system around the latter but not the former, and the twice higher enrichment of heavy elements in Saturn vs. Jupiter (Atreya et al., 2019).

The basic outlines of the core accretion model are as follows: A heavy element core grows through pebble, planetesimal and embryo accretion in the gaseous disk (pebbles are cm-to-meter sized bodies, embryos lunar and upward, and planetesimals in between). We may call this phase I after Helled et al. (2014). At some point, the gravitational attraction of the growing core enhances the accretion rate of the gas, while the increasing thermal energy content of the atmosphere slows the rate, so that the core and envelope grow at approximately the same, slow rate. This is phase II. Phase III is initiated when sufficient gas has accreted that its gravity overcomes the thermal pressure and there is a rapid, runaway collapse of gas onto the protoplanet. The truncation of phase III is not entirely clear but may be related to the formation of a deep gap in the surrounding disk, which at least slows the supply of gas.

Such a model produces, by definition, a heavy element core, and might be expected to also feature an envelope enriched in heavy elements, but the latter process is complicated by the interaction of the solids with the gas. The complexity of the interaction is tied to the solution to the long-standing problem of the lengthy time required to get through Phase II. Realization that cm- to meter-sized grains, or "pebbles", in the disk would have behaviors strongly affected by the gas (Johansen and Lambrechts, 2017) has led to a reworking of the core growth models which dramatically increase growth rates (Lambrechts et al 2014). However, pebbles are also much more sensitive to the presence of pressure bumps at the edges of the disk gaps caused by the growing giant planets (Bitsch, et al., 2018), such that accretion rates of gas vs. solids may vary in a complex fashion as planets transition from Phase II to Phase III and begin to induce gap formation. As a consequence, some or most of the heavy element enrichment might be contributed by molecules in the gas, rather than the solid phases in the disk, as quantified by Mousis et al. (2019, and Sec. 3.4 here). As they show, the decoupling of the pebble-sized grains from the accretion process during some or all of Phase III could alter the ratio of water to more volatile species in the envelope of Jupiter depending on where that planet formed. Since the formation of Saturn took place in a different location in the disk, it is possible that, for example, the ratio of oxygen to carbon in Saturn is different from that in Jupiter.

Regardless of the above complications, most of the overall features of both Jupiter and Saturn argue for core formation rather than the competing "disk instability" model (e.g., Boss 1997; Mayer et al. 2002, 2004). In the latter model, the disk is unstable to breakup, and forms gaseous concentrations on a very rapid timescale of hundreds of years to thousands of years. These condensations have the same metallicity as the gas, and presumably the central star, but will capture additional heavy elements into their envelopes as material condenses out of a second generation, stable, disk. That Jupiter and Saturn have cores of roughly the same mass (despite the different configurations) argues for a common formation process in which the core mass controls the rate of envelope accretion, and therefore the core preexists the envelope. In the disk instability model, the disk is too young, and the breakup occurs much too quickly, for a core to form first from some population of solid material (dust, pebbles, planetesimals, or embryos); the core must be the result of the accretion of enough material contained in large bodies that much of it can reach the center without completely dissolving in the envelope. Simulations suggest that such cores would be largely made of rock-forming elements (Helled and Schubert, 2008) which would put a lot of ice-forming elements into the envelope, possibly violating the Juno-based constraints for Jupiter (Wahl et al 2017). Although the diffuse core inferred for Jupiter might at first glance seem to argue for such a disk instability scenario of post collapse addition of core material, the disk instability model does not explain the existence of Uranus and Neptune.

The presence of a much larger enrichment of heavy elements in Saturn vs. Jupiter coupled with the comparable sizes of their cores tilts strongly in favor of the core accretion model for giant planet formation as opposed to the disk instability model since the latter has no natural explanation for the two heavy elements configurations. In the core accretion model the core sizes drive the envelope accretion rates in Phase II while much of the envelope heavy element inventory is added in late Phase II and Phase III. This is about as much as one can say without knowing the bulk oxygen abundance in Saturn, or the noble gas abundances. In Jupiter, the roughly uniform abundances of the noble gases and carbon (Niemann et al. 1998, Mahaffy et al. 1998 and 2000, Wong et al 2004), coupled with the limits on the water abundance from Juno atmospheric data (Li et al. 2020) and interior structure constraints (Wahl et al. 2017) argues for a particular model of heavy element seeding involving pebbles isolated at the Jovian pressure bump in the disk (Mousis et al 2019). We cannot know whether such a model applies to Saturn—or has even been correctly interpreted for Jupiter—without comparable Saturnian data sets. But if the interpretation of the Juno and Galileo probe data is correct, it represents an echo of key properties and processes (pebbles, gap formation) associated with Phase III of giant planet formation in the core accretion model.

Finally, that Saturn apparently has a diffuse core (Mankovich and Fuller, 2021) directs attention to the possibility that Jupiter and Saturn shared a common, and potentially complex, set of processes that infused heavy elements throughout their interiors. Liu et al. (2019) argued that the most plausible way to form a diffuse core is to start with a discrete core and have Jupiter suffer a mega-impact, from a rogue planetary core of about 10 $M_{Earth}$. However, it seems less plausible that both giant planets suffered the same fate, especially given that rather specific conditions are required for an impact to produce the desired results (Helled et al., 2022). Formation models that include dissolution of heavy element material in the envelope produce an envelope gradient (Helled and Stevenson, 2017; Valletta and Helled, 2020), but not sufficient to explain the details of Jupiter's interior (Helled et al 2022). Whether the same is true for Saturn is unclear given the data at hand.

The differences between Jupiter and Saturn are emphasized in the architecture of their satellite systems. Early attempts to understand why Jupiter has a regular satellite system of four large bodies arranged in order of ice content, and with roughly the same mass of rock (within a factor of two) in each, centered on the presence of a dense circumJovian disk of material. Such a disk is a natural outcome of the final gas accretion phase. This provides a natural explanation for the gradient in rock to ice, and potentially an explanation for the lack of geologic activity on Callisto in terms of an early version of the pebble accretion model (Lunine and Stevenson, 1982). However, the dense gaseous disk was considered problematic from several points of view, and gas-starved disk models replaced these (Canup and Ward, 2002). Saturn, in contrast, has only one large regular satellite, Titan, and many small ones. In neither case is the difference from Saturn explained, although models postulating one or more additional Titan-sized bodies, all but one of which were lost to Saturn (Canup 2010), suggest the difference between the two systems is not directly to do with the formation of the planets themselves. Such a scenario does potentially explain the existence of Saturn's massive ring system and its cohort of smaller satellites (which would be born from the rings, Charnoz et al. 2011, Crida & Charnoz 2012), and their absence from Jupiter, if the rings are ancient rather than young (still controversial). However, differences in the amount of circumplanetary material might have been responsible (Mosqueira and Estrada, 2003). How such differences relate to the formation of Jupiter and Saturn themselves is unclear. More recent models focusing on the role of pebble accretion (Shibaike et al 2019) and the possibility of migration of Saturn into resonance with Jupiter injecting large amounts of circumsaturnian material into the Jupiter gravity well (Ronnet et al., 2018) may lead to new possibilities that the architectures of the two systems were influenced by gravitational interactions between the two giant planets. Also, the location of the centrifugal radius is key in circum-planetary disks (CPDs) to assess the formation conditions of the moons because so far, many models assume the injection point of matter is confounded with this location. For example, Anderson et al. (2021) find that assuming classically a centrifugal radius in the 20– 30 $R_{Sat}$ range (Ruskol 1982; Canup & Ward 2002, 2006) makes the condensation of volatiles-rich pebbles extremely difficult, except for those condensed from water. Had Enceladus and Titan formed from solids in the CPD, this result would be at odds with our current knowledge of their compositions. On the other hand, the authors find that simulations of pebbles transport and condensation in a Saturn's CPD with a centrifugal radius in the 66–100 $R_{Sat}$ range allow for the formation and growth of solids with compositions consistent with those measured in Enceladus and Titan, but Enceladus is unlikely to be primordial anyway (also Sec. 3.5).

## 3.3 Atmospheric Composition Constraints

### *3.3.1 Helium*

Elemental abundances and their isotopic ratios provide by far the best chemical constraints to the formation and evolution models of Saturn in particular, and giant planets in general. The noble gases, He, Ne, Ar, Kr and Xe are chemically inert, hence unaffected by meteorology and dynamics. Unlike Jupiter, the only noble gas measured in Saturn is helium. Whereas Jupiter's He could be measured with high precision *in situ* by the Galileo probe, Saturn's helium is derived indirectly by remote sensing observations, and the results are puzzling. Helium is key to understanding Saturn's evolution as the planet cools and contracts over time.

#### *Observations*

Several new results on Saturn's helium abundance have become available because of recent Cassini observations. All available He abundances to date for Saturn along with a comparison to the other giant planets and the Sun are listed in Table 3.1 and plotted in Figure 3.2. Pre-Cassini results on $He/H_2$ volume fraction ranged from as low as 3.5% to as high as 13.5% by volume. The low value is less than ¼ of Jupiter's $He/H_2$, which is 15.7%. The analysis from Cassini using the UVIS and IR data yields a value of 12.4% by volume (Koskinen et al., 2018), but it is model dependent, subject to assumptions of extrapolation of mixing from above the poorly constrained homopause level – supposedly located between $10^{-6}$ and $10^{-8}$ bar – to the well-mixed troposphere. The latest analysis of the Cassini CIRS data yields a much lower value of 0.04-0.075, or an average 5.75% by volume (Achterberg and Flasar 2020), which is consistent with a value of 5.5% based on modeling of VIMS spectra of the cloud clearing during the aftermath of Saturn's Great Storm of 2010-2011 (Sromovsky et al. 2016). On the other hand, measurements made by the Cassini INMS in the Grand Finale orbits before the spacecraft entered Saturn yield a $He/H_2$ ratio of 0.10-0.16. The value at the higher range (0.16) is nearly identical to the $He/H_2$ ratio at Jupiter from the Galileo Probe, while the lower number (0.10) accounts for additional helium rain in the interior of Saturn compared to Jupiter. However, it should be noted that the deepest INMS data correspond to is ~1700 km above the 1-bar level, which is several hundred kilometers above the homopause, thus the $He/H_2$ ratio in the well-mixed atmosphere remains model dependent, like the abovementioned combined UVIS+IR result. The bewildering range of presently available $He/H_2$ ratio in Saturn's atmosphere argues strongly for a direct in situ measurement of this fundamental quantity from an entry probe. Until then it is premature to favor one particular value over the other and it would be prudent to assume a plausible range of 3-13% for the helium mole fraction at Saturn based on available data.

Table 3.1. *Helium Fraction in the Sun and the Atmospheres of the Giant Planets*

| Object | He<br>Mole fraction | $He/H_2$<br>Volume Fraction | $He/H_2$<br>Mass Fraction [l] |
|---|---|---|---|
| Protosolar[a] | — | 0.191 | 0.276 |
| Jupiter[b] | 0.136±0.003 | 0.157±0.003 | 0.238±0.003 |
| Saturn | | | |
| *Cassini* | | | |
| INMS[c] | 0.095-0.137 | 0.100-0.160 | 0.167-0.242 |
| CIRS[d] | 0.038-0.071 | 0.0575±0.0175 | 0.102±0.028 |
| CIRS+UVIS[e] | 0.110±0.020 | 0.124±0.025 | 0.199±0.031 |
| VIMS[f] | $0.052^{+0.009}_{-0.014}$ | $0.055^{+0.010}_{-0.015}$ | $0.099^{+0.016}_{-0.025}$ |
| *Voyager* | | | |
| IRIS/RSS[g] | 0.033±0.027 | 0.034±0.028 | 0.064±0.055 |
| IRIS only[h] | 0.118±0.023 | 0.135±0.025 | 0.212±0.031 |
| UVS[i] | 0.131±0.020 | 0.150±0.025 | 0.226±0.030 |
| *Pioneer 10*<br>IRR/RSS[j] | 0.099±0.027 | 0.110±0.030 | 0.178±0.040 |
| Uranus[k] | 0.150±0.025 | 0.180±0.030 | 0.265±0.031 |
| Neptune[k] | | | |
| without $N_2$ | 0.190±0.030 | 0.240±0.038 | 0.324±0.036 |
| with 0.3% $N_2$ | 0.150±0.025 | 0.180±0.030 | 0.265±0.031 |

[a]Asplund et al. (2009); [b]von Zahn et al. (1998); [c]Waite et al. (2018, see text);
[d]Achterberg and Flasar (2020) - note that these authors present $He/H_2$ as a range of 0.040-0.075, which we have expressed as 0.0575±0.0175 in the above table plotted in the corresponding figure;
[e]Koskinen and Guerlet (2018); [f]Sromovsky et al. (2016); [g]Conrath et al. (1984);
[h]Conrath and Gautier (2000); [i]Ben-Jaffel and Abbes (2015), derived from He 584 Å emission above the homopause and highly dependent on the assumptions of radiative transfer, atmospheric mixing and thermal structure; [j]Orton and Ingersoll (1980);
[k]Gautier et al. (1995);
[l]Conversion of helium volume fraction to mass fraction:
$\frac{M_{He}}{M_{H_2}+M_{He}} = \frac{4[He/H_2]}{2[H_2/H_2]+4[He/H_2]}$, where $M_x$ is the molecular mass of gas x and $He/H_2$ is the volume fraction of He to $H_2$.
CIRS: composite infrared spectrometer, UVIS: ultraviolet imaging spectrograph,
VIMS: visual and infrared mapping spectrometer, IRR; infrared radiometer,
RSS: radio science subsystem, IRIS: infrared interferometer spectrometer.

### *Models and Experiments*

Models had previously suggested that in the deep interiors of Jupiter and Saturn, helium becomes immiscible with hydrogen at megabar pressure levels (helium becomes immiscible with metallic hydrogen, Stevenson and Salpeter, 1977, Stevenson, 1985). Subsequent *ab initio* thermodynamic calculation for hydrogen-helium mixtures also predicts such a phase separation in Saturn's interior (Schöttler and Redmer 2018, 2019). The thermal evolution model of Mankovich and Fortney (2020) predicts a much more severe differentiation of helium in Saturn's interior than Jupiter, resulting in a helium-rich shell or core and a greatly depleted ratio of $He/H_2=0.036$ in the atmosphere, which agrees with the very low end of the values discussed in the previous section. Though the interior models have long predicted hydrogen-helium demixing, simulating the high p-T conditions in the laboratory has proved to be quite challenging. Brygoo et al. (2021) have now successfully carried out laser-driven shock compression experiments, which provide evidence of helium-hydrogen immiscibility at critical temperatures ranging from 4700 K at 93 GPa to 10,200 K at 150 GPa, implying that the helium-rain region may cover a relatively large fraction of the interiors of Saturn and Jupiter.

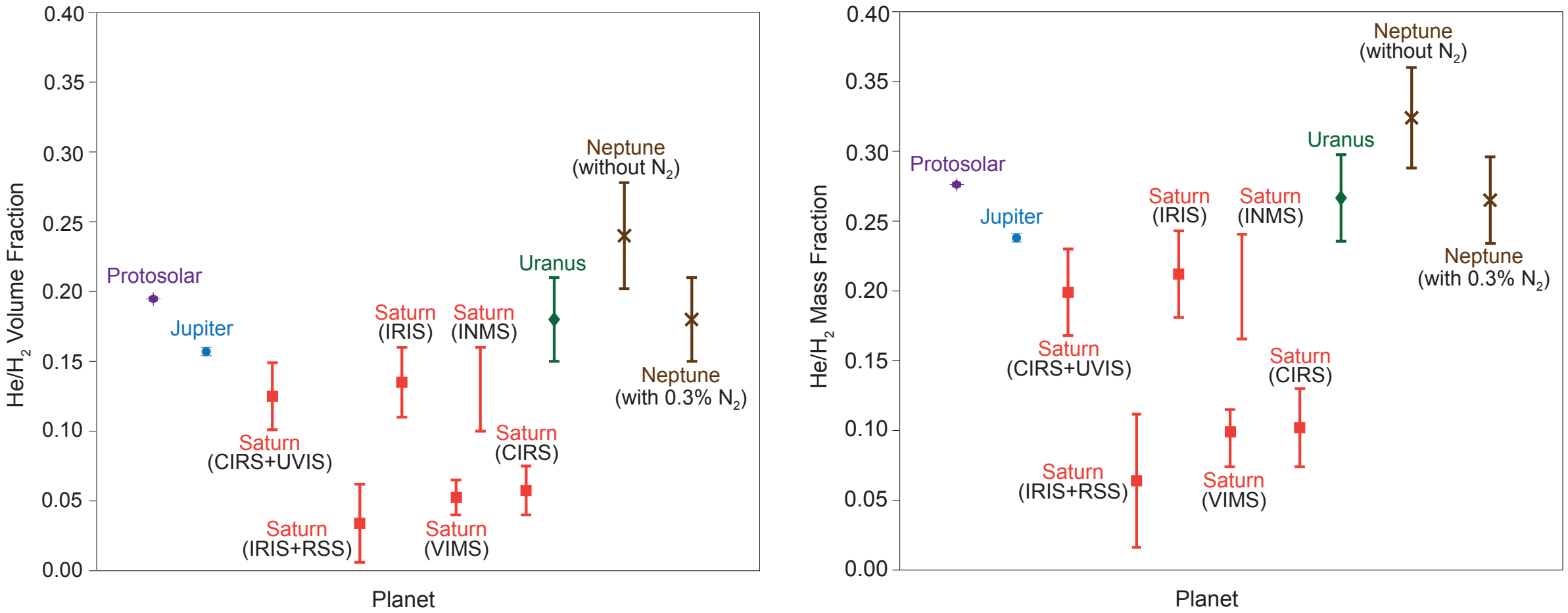

Figure 3.2 The $He/H_2$ ratio in the atmosphere of Saturn by volume and mass, compared to the Sun and the other giant planets. Saturn's values reflect a wide range depending on the instruments and methods employed for retrieval. See Table 3.1 and associated footnotes for the values, explanations, and references.

### *Helium Rain, Heat Balance. Luminosity, and Saturn's Evolution*

Both Saturn and Jupiter have excess luminosity that cannot be attributed to their primordial heat of formation as the planets cool and contract gradually over time. It has been known since the Voyager flybys between 1978 and 1980 that both planets emit nearly twice as much energy as they absorb from the Sun (Conrath et al., 1984; Li et al. 2018), implying a large source of internal energy. [Approximately 1.0002 for the Earth.] Planetary heat flux is a fundamental quantity for understanding the evolution of Saturn and Jupiter, as well as convective processes in their tropospheres. The heat of formation is a part of the observed heat flux. The excess luminosity of Jupiter is most likely due to abovementioned demixing of helium from metallic hydrogen. The nearly 20% depletion of helium below solar in Jupiter's atmosphere measured by the Galileo probe (Table 3.1) is direct evidence of helium condensation in Jupiter's interior that is further bolstered by the finding of greatly sub-solar atmospheric neon, as Ne is removed along with helium into which it dissolves (Roulston and Stevenson, 1995; Wilson and Militzer, 2010). As helium rain drops separate and fall to the core, their gravitational potential energy is converted to heat, which contributes to Jupiter's luminosity. On the other hand, Saturn is smaller and cooler than Jupiter, yet it has a very large excess luminosity, which requires an even greater degree of helium differentiation in Saturn's interior. The atmospheric $He/H_2$ ratios at the lower range of available data (Table 3.1) support this idea, as do theoretical models (previous section). In the absence of this process, Saturn's present luminosity would have been reached in roughly half the age of the solar system (Grossman et al., 1980; Schöttler and Redmer 2018).

#### *Large $He/H_2$ Ratio at Saturn?*

The measured luminosity of Saturn argues strongly for gravitational settling in the interior and therefore a low helium abundance (Fortney and Hubbard 2004, Mankovich and Fortney 2020), as discussed above. However, there exist a few results, which, at the upper range of their uncertainty, indicate values as high as Jupiter's $He/H_2$ (Table 3.1). Even more puzzling result would be solar or super-solar $He/H_2$, especially viewed in the context of the slightly sub-solar value for Jupiter, and the fact that even though photoevaporation in the disk may be invoked, it is not expected to favor the ejection of helium over hydrogen (Guillot and Hueso 2006). One possibility to accommodate the high value of helium in Saturn's atmosphere, if correct, and the planet's large luminosity may be through the presence of large compositional gradients in the interior, leading to a slow leakage of primordial heat (Leconte and Chabrier 2013). Resolving this issue is hence crucial to really understand Saturn's internal structure and evolution.

### *3.3.2 Nitrogen and Oxygen (Ammonia and Water)*

Ammonia and water are very important components of Saturn's atmosphere, but because they condense as clouds within or below the observable part of the atmosphere, their deep (bulk) abundance is particularly difficult to assess. Depending on its concentration, water vapor condenses in Saturn's atmosphere down to approximately 20 bars, too deep for a direct estimate of its abundance (Atreya et al. 2019). Ammonia on the other hand condenses at pressures of near 2 bars (Atreya et al. 2019, Fig. 9) and its infrared bands may be detected by spectroscopy. Using Cassini/VIMS data between 4.6 and 5.1 μm, Fletcher et al. (2011) infer a latitude-dependent ammonia mixing ratio of 300-500 ppmv in a ±5° latitude band centered on the equator and between 100 and 200 ppmv at most other latitudes. The situation is like that in Jupiter's atmosphere seen with Juno's microwave radiometer (MWR), where ammonia in the 1-3 bar region (below its condensation level at 0.7 bar) also shows an elevated abundance, around 350 ppmv (Li et al. 2020) in the equatorial zone, with lower values generally around 100-200 ppmv in other regions. In Jupiter's deeper atmosphere, at pressures greater than approximately 30 bars, the $NH_3$ abundance appears to recover to a uniform value of 351±22 ppmv, independently of latitude (Li et al. 2017, 2020).

The Jupiter observations imply that some mechanism is transporting ammonia efficiently from the upper levels down to deeper pressures and in such a way that it is essentially invisible to the global MWR observations (Ingersoll et al. 2017). Guillot et al. (2020a) propose that convective storms driven by the latent heat of water (and to a lesser extent, ammonia) condensation lead to the formation of large ammonia-water hailstones ("mushballs") that fall rapidly and survive to below the base of the water cloud (5-8 bars). Upon evaporation, these cool, high-molecular weight regions should produce strong downdrafts and transport ammonia and water further down into the deep atmosphere. A mass-exchange model shows that explaining the MWR observations requires a latitude-dependent storm rate in Jupiter that correlates with the measured lightning flash rate at least in the ±20° latitude region, and a deep atmosphere that is stable on average (Guillot et al. 2020b).

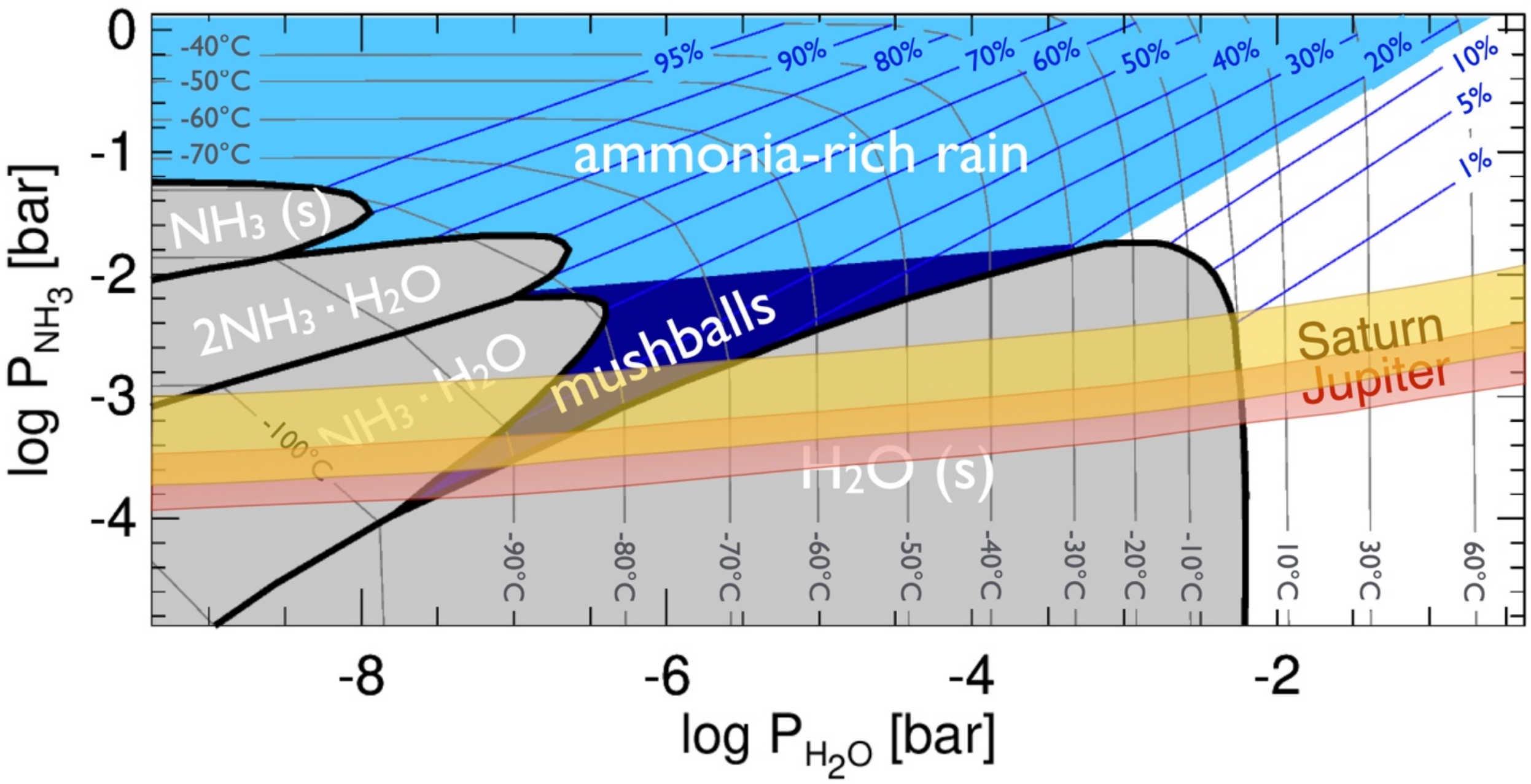

Figure 3.3 $H_2O$-$NH_3$ equilibrium phase diagram (Weidenschilling & Lewis, 1973) as a function of partial pressure of $H_2O$ and $NH_3$. Solid phases are indicated in gray, otherwise, a liquid mixture forms with a concentration in ammonia indicated by the blue diagonal contour. The temperatures in celsius are indicated as contour lines running from the bottom to the left of the plot. The red region labeled Jupiter spans conditions in Jupiter's atmosphere assuming a minimum $NH_3$ abundance of 100 ppmv and a maximum value of 360 ppmv (Li et al., 2017). The yellow region labeled Saturn spans conditions in Saturn's atmosphere assuming an $NH_3$ abundance between 100 and 500 ppmv (Fletcher et al., 2011). [Adapted from Guillot et al. 2020b].

In Saturn, condensation models predict that the region of formation of the ammonia-water liquid at the origin of mushballs should be more extended. Figure 3.3 shows that while the mushball formation region is limited to 1.1 to 1.5 bar in Jupiter, corresponding to a 9 km vertical extent, it ranges from 2 to 3.2 bar in Saturn, a 35 km vertical extent. A key question is whether water storms can loft a large-enough amount of ice crystals from approximately 10 bar region up to at least the 3.2 bar level. Given the properties of storms in Saturn (e.g., Li & Ingersoll 2015), this is likely. Spectroscopic evidence for water ice in the Great Storm of 2010–2011 also suggests that at least in the largest of storms, vertical velocities are sufficient (Sromovsky et al. 2013).

A natural interpretation of the latitude dependency of the ammonia abundance in Saturn is thus that, as Jupiter, for a yet unknown reason (see Guillot et al. 2020b), strong storms are rare or absent in the equatorial zone which is mixed by diffusion and/or small-scale convection and maintains a high ammonia abundance. Other regions are dominated by storm activity, leading to the formation of mushballs and ammonia-water rich downdrafts, and therefore a low ammonia abundance at the high altitudes probed by infrared spectroscopy. If, as seems to be the case on Jupiter, the equatorial zone is uniformly mixed with the deeper region, Saturn should have a bulk abundance of $NH_3$ between 300 and 500 ppmv. It is possible however that less frequent storms also create a gradient in the equatorial zone in which case the bulk $NH_3$ abundance may be higher than these values.

Saturn's water abundance is presently unknown. However, modeling of Saturn's giant storms observed by Cassini can provide important constraints on its possible range of values. An indirect method of inferring the atmospheric composition is through its effect on the atmospheric dynamics. Saturn emits 78% more heat than it receives from the Sun (Hanel et al. 1983), meaning that the internal heat accumulated during the formation of the planet is carried out by convection until it reaches the level where the gas opacity is low enough so that the photons can escape into space. The standard theory of convection, mostly formulated in a Boussinesq framework, predicts that the thermal structure is nearly adiabatic in the body of the fluid. However, the presence of "heavy" molecules such as water and ammonia in a background gas made of "light" molecules such as hydrogen and helium complicates the standard picture, as discussed above.

Condensation of water has two consequences. First, the release of latent heat warms the background atmosphere. The moist adiabatic temperature gradient – temperature gradient resulting from an adiabatically lifted moist air parcel while water is condensing along the way – is less than the dry counterpart because additional heat is released to offset the drop of temperature due to adiabatic expansion. The temperature difference between the moist and dry adiabatic expansion is approximately $xL/c_p$, where x is the water vapor mixing ratio, L is the latent heat of condensation and $c_p$ is the weighted average heat capacity of the background atmosphere. The more water the air parcel contains, higher the temperature it can attain when all the water vapor has condensed.

On the other hand, condensation removes water vapor from the atmosphere, causing a sharp reduction in the mean molecular weight above the condensation level. Convection across a compositional gradient is poorly studied. Limited works on linear stability analysis (Guillot 1995; Friedson and Gonzales, 2017; Leconte et al., 2017) and two-dimensional numerical modeling (Sugiyama et al. 2014; Li and Chen 2019) demonstrate that the temperature gradient may be super-adiabatic since the reduction of mean molecular weight allows a cold and dry air parcel to have the same density as a warm but moist one. If the water abundance exceeds 10 times solar, the molecular weight effect overwhelms the latent heat effect such that convection occurs intermittently. Conversely, if the water abundance is less than 10 times solar, the latent heat effect dominates, and convection occurs continuously. Observations of giant storm eruptions on Saturn with periodicities of ~30 years (Sánchez-Lavega et al. 2012) suggest that Saturn has more than 10 times the solar abundance of water while Jupiter has less than 10 times solar (Li and Ingersoll 2015). Water abundances in Jupiter and Saturn from analysis of storm associated lightning are consistent with these values (Dyudina et al. 2010, 2013). The recent Juno MWR observation tends to support this argument by measuring the water abundance at the equatorial zone to be between 1.5 and 8.3 times solar (Li et al., 2020). Ongoing analysis of the Juno MWR data is expected to determine the *global* abundance of water at Jupiter as well as its depth profile, whether there actually is a gradient in its mixing ratio below the lifting condensation level creating a region that is on average stable and may be superadiabatic as surmised above. Extrapolation to Saturn would then provide better insight into the connection between Saturn's giant storms and the planet's water abundance.

The existing atmospheric models for Jovian planets are still too simplistic to represent the complicated interaction between dynamics and thermodynamics. The Juno mission found that ammonia gas on Jupiter only reaches its well-mixed abundance at very deep levels (Bolton et al., 2017; Li et al., 2017), so that the ammonia concentration at the cloud base is lower than in equilibrium condensation models. Therefore, inferring the deep O/H abundance based by its effect on the dynamics requires elaborate atmospheric models.

### *3.3.3 CO and other Disequilibrium Species*

While remote sensing and in situ observations are confined to relatively shallow depths, disequilibrium species can give an insight into the deep atmospheric convective processes, and thereby help with the interpretation of certain heavy element abundances. In Saturn's observable upper troposphere, the chemical equilibrium reservoirs of C and O are $CH_4$ and $H_2O$. But chemical equilibrium only holds at depths greater than the "quench level," where the timescale for thermochemical equilibration equals the vertical mixing timescale (Figure 3.4). Above the quench level, turbulent diffusion can mix disequilibrium species including CO, and potentially $CO_2$ and $C_2H_6$, into the upper levels of the troposphere. Additional disequilibrium species observed in Saturn's troposphere include $PH_3$, $GeH_4$, and $AsH_3$. We summarize disequilibrium species concentrations in Saturn's troposphere in Table 3.2. Some values are averaged from multiple measurements that still contain significant discrepancies, in which cases error bars have been expanded beyond published estimates to encompass the differences in the results.

Disequilibrium species are affected by photochemistry at/above the tropopause (Fig. 3.4), but they are not affected by tropospheric cloud condensation. This makes them particularly valuable tracers of sub-cloud properties such as elemental abundance ratios and the rate of mixing, parameterized by the vertical eddy diffusion coefficient $K_{zz}$ (e.g., Fegley and Lodders 1994). Modeling of the relevant chemical kinetics and turbulent diffusion of these species is an active area of research, where large discrepancies and uncertainties exist between reaction schemes and rates, turbulent mixing rates, and upper tropospheric abundance retrievals themselves.

Using the tropospheric CO abundance to constrain Saturn's O/H ratio is particularly appealing because $H_2O$ itself condenses near 27 bars assuming 21× solar O/H (Fig. 3.4), and water may be depleted (with respect to its deep abundance) even deeper below the cloud base if convective processes operate similarly on Jupiter and Saturn (Li et al. 2017, Guillot et al. 2020). Methane does not condense in Saturn's troposphere, so the C/H ratio measured by remote sensing (e.g., Courtin et al. 1984, Flasar et al. 2005, Fletcher et al. 2009) can be fixed in thermochemical models. The disequilibrium species $CO_2$ is predicted to be ~10% of the CO concentration for significant super-solar enrichments of water (e.g., O/H = 21× solar in Fig. 3.4). Measurements of $CO_2$ exist for Saturn's stratosphere (Moses et al. 2000, Abbas et al. 2013), where external flux from the rings and micrometeorites supply oxygen species (Moses and Poppe 2017), but there are no measurements in the troposphere. A new estimate using CO in this manner gives an O abundance 7-15 times the protosolar value (Cavalié et al. 2024). This is consistent with a water abundance of of 8 times solar derived from lightning models constrained by Cassini data (Aglyamov et al. 2023).

Phosphine ($PH_3$) is another disequilibrium species, affected by photochemistry but not cloud condensation. Phosphine is an appealing observational target because of its distinct spectral signatures in the thermal infrared. Its tropospheric concentration may be sensitive to the O/H ratio as well as the P/H ratio, due to thermochemical equilibrium with species such as $P_4O_6$ or $H_3PO_4$ (Fegley and Lodders 1994; Wang et al. 2015, 2016).

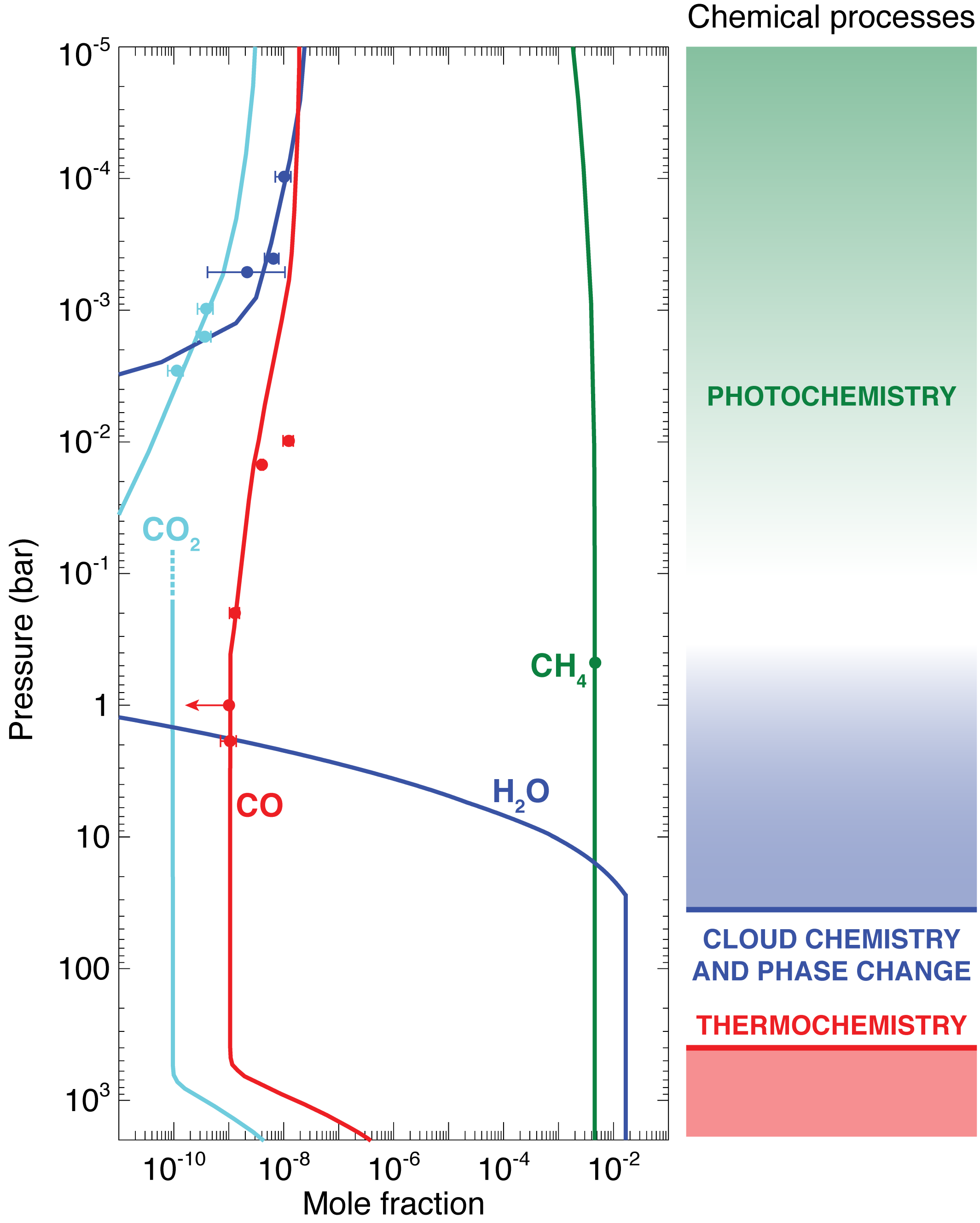

Figure 3.4 In the deep atmosphere, thermochemistry partitions elements such as O and C between several molecules such as CO, $CO_2$, $H_2O$, and $CH_4$ (red region). Their abundance ratios are frozen—no longer subject to thermochemical equilibrium—above "quench levels," unique to each species, that ultimately depend on abundances, thermal profiles, and turbulent diffusion rates. Disequilibrium species should have constant abundances in the deep troposphere, but their concentrations are also affected by chemical processes such as photochemistry in the stratosphere (green) and condensation in the troposphere (blue). Saturn profiles

shown here are from a thermochemical model in the deep troposphere (Wang et al. 2016), a cloud condensation model in the upper troposphere (Wong et al. 2015), and a photochemical model with dust influx in the stratosphere (Moses and Poppe 2017). Profiles for $CO_2$ are disjoint because Moses and Poppe (2017) did not consider an internal source. Data points give spectroscopic measurements of $CO_2$ (Moses et al. 2000, Abbas et al. 2013), $H_2O$ (Bergin et al. 2000, Moses et al. 2000, Fletcher et al. 2012), CO (Noll and Larson 1990, Moses et al. 2000, Cavalié et al. 2009; 2010), and $CH_4$ (Courtin et al. 1984); see Moses and Poppe 2017 for more information on the measurements).

Table 3.2 *Disequilibrium Species in the Tropospheres of the Gas Giant Planets*

| Mole fraction | Jupiter | Saturn |
|---|---|---|
| CO | $9 \pm 3 \times 10^{-10}$ [a,b] | $1.0 \pm 0.3 \times 10^{-9}$ [k] |
| $PH_3$ | $1.35 \pm 0.6 \times 10^{-6}$ [a,c] | $5.2 \pm 1.6 \times 10^{-6}$ [c,l] |
| $GeH_4$ | $6 \pm 3 \times 10^{-10}$ [a,g] | $4 \pm 4 \times 10^{-10}$ [m] |
| $AsH_3$ | $4.5 \pm 1.9 \times 10^{-10}$ [e,f] | $3 \pm 1 \times 10^{-9}$ [k] |
| $SiH_4$ | $< 2.5 \times 10^{-9}$ [d] | $< 2 \times 10^{-10}$ [k] |

[a]Bjoraker et al. (2018); [b]Bézard et al. (2002); [c]Fletcher et al. (2009); [d]Treffers et al. (1978); [e]Giles et al. (2017); [f]Noll et al. (1990); [g]Grassi et al. (2020); [k]Noll and Larson (1991); [l]Fletcher et al. (2011); [m]Noll et al. (1988)

Tropospheric abundances of $GeH_4$ and $AsH_3$ (along with $SiH_4$, which has never been measured), may someday offer a chance to determine rock/ice ratios for solids accreted during giant planet formation, after O/H and disequilibrium species abundances are well determined.

Chemical-diffusion models that simultaneously match observations of multiple disequilibrium species at once offer the best chance to break degeneracies between deep abundances and mixing rates, especially because eddy mixing rates themselves may be variable in both the vertical and horizontal directions (Wang et al. 2015). Improvements in chemical reaction rates are needed. Some progress also remains to be made reconciling spectroscopic measurements at multiple wavelengths for $PH_3$; 10-µm retrievals give consistently higher values than 5-µm retrievals (Fletcher et al. 2009; 2011, Giles et al. 2017, Bjoraker et al. 2018). For $AsH_3$ on Jupiter, models predicting almost no sensitivity to $K_{zz}$ (Wang et al. 2016) are difficult to reconcile with a more than 3× increase from equatorial regions to high latitudes reported from both high-resolution ground-based spectroscopy and lower-resolution spectroscopy on the Juno polar orbiter (Giles et al. 2017, Grassi et al. 2020).

### *3.3.4 Summary of Heavy Elements and Helium and an Update on the Xenon Isotopes*

Presently known abundances of the elements critical to Saturn's formation models are listed in Table 3.3 and illustrated in Figure 3.5. Additional details are contained in preceding sections. Compared to Jupiter, little information on elemental and isotopic abundances currently exists for Saturn. Most of the available data comes from Cassini remote sensing observations. While Cassini/CIRS has determined C/H precisely, N/H is tentative since it is derived from $NH_3$ measured only down to depths just below its condensation level but $NH_3$ could be higher in the deeper atmosphere considering that the well-mixed region of $NH_3$ Juno discovered at Jupiter is tens of bars below the ammonia clouds. O/H is unknown, but estimates based on dynamical modeling of Saturn's storms and associated lightning place it roughly in the same range as C/H. S/H is highly questionable, considering the challenges associated with its model-dependent retrieval from the VLA data (see Atreya et al. 2019 for additional details). P/H is listed for completeness, but caution should be exercised in using it as a constraint on the formation models as it could be higher in the deep atmosphere, since it is derived from measurements of $PH_3$ in the shallow observable part of the troposphere where the gas is in disequilibrium, reaching thermochemical equilibrium much deeper at ~1000 K, 1000 bar level (Sec. 3.3.3). On the other hand, $PH_3$ and other disequilibrium gases, $GeH_4$, $AsH_3$ and CO, are indicators of strong convection from deep atmosphere. Noble gases are a critical input to the models of the formation and evolution (see, e.g., Atreya et al. 2019), but only the $He/H_2$ ratio is available for Saturn and it is highly uncertain. The Galileo Probe provided an excellent set of data on the abundances of the noble gases and their isotopic ratios at Jupiter. Formation models predict similar values at Saturn, but only an entry probe can confirm that. It would be important to measure the noble gases with high precision in order to discriminate between competing formation models. However, there is some confusion about the precision of xenon isotope measurements made by the Galileo Probe in Jupiter's atmosphere. The uncertainties listed in table 1 of Mahaffy et al. (2000) are seemingly too large to be meaningful. On the other hand, using the same listing, Atreya et al. (2019 - table 2.2 of Chapter 2 of the previous version of this book) reported very small uncertainties. This apparent discrepancy is due to the way uncertainties were reported in the two papers. While Mahaffy et al. normalized the uncertainties of individual xenon isotope to total xenon ratios by the corresponding solar ratios, Atreya et al. eliminated the normalization factor (by multiplying Mahaffy et al.'s uncertainties by the corresponding solar ratio). Thus, the uncertainties listed in Atreya et al. (2019) represent the actual error bars and they show that the Galileo Probe achieved a very high degree of precision in the determination of xenon isotope ratios at Jupiter. This is further confirmed by the nearly identical standard deviation calculated from the set of xenon measurements made by the mass spectrometer in three parts of the Galileo Probe descent, where significant xenon counts were recorded (P. R. Mahaffy, personal comm. 2024). The values with standard deviation are: $^{136}Xe/Xe=0.0805\pm0.00981$, $^{134}Xe/Xe=0.0956\pm0.00686$, $^{132}Xe/Xe=0.2881\pm0.01612$, $^{131}Xe/Xe= 0.2015\pm0.01630$, $^{130}Xe/Xe=0.0416\pm0.00634$, $^{129}Xe/Xe=0.2748\pm0.01576$, $^{128}Xe/Xe= 0.0179\pm0.00057$. Like xenon, other noble gas isotopes were also determinized at high precision by the Galileo Probe ($^3He/^4He=1.66\pm0.05\times10^{-4}$, $^{36}Ar/^{38}Ar=5.6\pm0.25$, and $^{20}Ne/^{22}Ne=13\pm2$). Based on the above measurements at Jupiter, one can expect to determine the abundances and isotopic ratios of noble gases at similar or higher precision on a future entry probe at Saturn.

Table 3.3 *Elemental Abundances in the Sun and the Atmospheres of the Giant Planets*

| Elements | Sun-Protosolar[a] | Jupiter[b] | Saturn[b,c] | Jupiter/Protosolar | Saturn/Protosolar | Uranus/Protosolar[d] | Neptune/Protosolar[d] |
|---|---|---|---|---|---|---|---|
| He/H | $9.55\times10^{-2}$ | $7.86\pm0.16\times10^{-2}$ | $2.88\pm0.88\times10^{-2}$<br>(?) *footnote (c)* | $0.82\pm0.02$ | $0.30\pm0.092$ | $0.94\pm0.16$[e] | $1.26\pm0.21$[e]<br>$0.94\pm0.16$[e] |
| Ne/H | $9.33\times10^{-5}$ | $1.24\pm0.014\times10^{-5}$ | NA | $0.13\pm0.001$ | NA | NA | NA |
| Ar/H | $2.75\times10^{-6}$ | $9.10\pm1.80\times10^{-6}$ | NA | $3.31\pm0.66$ | NA | NA | NA |
| Kr/H | $1.95\times10^{-9}$ | $4.65\pm0.85\times10^{-9}$ | NA | $2.38\pm0.44$ | NA | NA | NA |
| Xe/H | $1.91\times10^{-10}$ | $4.45\pm0.85\times10^{-10}$ | NA | $2.34\pm0.45$ | NA | NA | NA |
| C/H | $2.95\times10^{-4}$ | $1.19\pm0.29\times10^{-3}$ | $2.50\pm0.10\times10^{-3}$ | $4.02\pm0.98$ | $8.98\pm0.34$ | $\geq80\pm20$[f] | $\geq80\pm20$[g] |
| N/H | $7.41\times10^{-5}$ | $2.04\pm0.13\times10^{-4}$[h]<br>$3.32\pm1.27\times10^{-4}$[k] | $2.12\pm0.53\times10^{-4}$<br>(?) *footnote (c)* | $2.76\pm0.17$<br>$4.48\pm1.71$ | $\geq2.85\pm0.71$ | $1.4^{+0.5}_{-0.3}$ [i]<br>$\geq10^{-2}$-$10^{-3}$[j] | $\geq10^{-2}$-$10^{-3}$[j] |
| O/H | $5.37\times10^{-4}$ | $2.23\pm1.35\times10^{-3}$[l]<br>$2.45\pm0.80\times10^{-4}$[k] | $3.76$-$8.05\times10^{-3}$[o] | $4.1\pm2.5$<br>$0.46\pm0.15$ | 7-15 | NA | NA |
| S/H | $1.45\times10^{-5}$ | $4.45\pm1.05\times10^{-5}$ | $1.88\times10^{-4}$<br>(?) *footnote (c)* | $3.08\pm0.73$ | 13.01 (?) | $37^{+13}_{-6}$ [i]<br>$\gtrsim0.4$-$1.0$[m] | $\gtrsim0.1$-$0.4$[n] |
| P/H | $2.82\times10^{-7}$ | $1.08\pm0.06\times10^{-6}$ | $\geq3.64\pm0.24\times10^{-6}$ | $3.83\pm0.21$ | $\geq12.91\pm0.85$ | NA | NA |

NA: Not available; ?: uncertain or controversial (see footnote/text).

[(a)]Protosolar values are based on the present day solar elemental abundances in the photosphere listed in table 1 of Asplund et al. (2009), increased by +0.04 dex, i.e. ~11%, with an uncertainty of ±0.01 dex, except for a slightly larger adjustment of +0.05 dex (±0.01) for He, to account for the effects of diffusion at the bottom of the Sun's convective zone on the chemical composition of the photosphere, as well as the effects of gravitational settling and radiative accelerations.

[(b)]Atreya et al. (2019, 2020) and references therein.

[(c)]Saturn's helium listed in this table is based on current analysis of Cassini CIRS data (Table 3.1), which is indirectly derived, but the full set of all available data − also indirectly derived and model dependent − span a wide range of values with $He/H_2$ as low as 0.034±0.028 to as high as 0.135±0.025 as shown in Table 3.1, corresponding to a range of Saturn/Protosolar ratio from 0.0314 to 0.8377, using the lower range of the lowest value and the upper range of the highest value. C/H is calculated using the central value for $He/H_2$=0.0575 from the latest CIRS analysis (Table 3.1) hence it is very slightly smaller than the value in Atreya et al. (2019, 2020). N/H corresponds to the maximum $NH_3$ mixing ratio of 400±100 ppmv measured down to 3 bars only (Fletcher et al. 2011), but greater values in the deep well-mixed atmosphere are entirely plausible; the Saturn/Protosolar N/H in the above table employs $NH_3$≥400±100 ppmv and the central value of $He/H_2$=0.0575 from CIRS analysis, while $He/H_2$=0.135 would give a ratio of ≥3.06±0.77, for example. S/H is based on a controversial interpretation of the VLA microwave data, hence dubious (see Atreya et al. 2019 for additional details). P/H is based on upper tropospheric abundance of $PH_3$ (Fletcher et al. 2009) but the actual P/H may be greater because $PH_3$ reaches thermochemical equilibrium only in the deep atmosphere (~1000 K, ~1000 bars) and its presence in the upper troposphere is due to strong upward mixing in the planet's interior (see Sec. 3.3.3).

[(d)]With the exception of He/H, elemental abundances for Uranus and Neptune should be considered as *lower limits*, not necessarily representative of their deep well-mixed atmosphere values.

[(e)]Gautier et al. (1995), for Neptune's He/H two values are listed, one without including any $N_2$ in the atmosphere (larger He/H) and the other including $N_2$ to explain the detection of HCL.

[(f)]Sromovsky et al. (2011) and personal communication (2015) with E. Karkoschka and K. Baines

[(g)]Karkoschka and Tomasko (2011), note that the C/H listed for both Neptune and Uranus is derived from $CH_4$ measurements in shallow troposphere to ~1 bar level, the region where CH4 condenses on these planets, hence the actual C/H in the deep well-mixed atmosphere could be greater.

[(h)]Juno microwave radiometer (Li et al. 2020, 2017; Bolton et al. 2017).

[(i)]Highest value derived for the deep atmosphere down at 50 bars near midlatitudes but it is found to be much less elsewhere, based on modeling of the ALMA data (Molter et al. 2021). Although deeper than the value at ~1 bar, 50 bars is still relatively shallow for the cold icy giant planets (Atreya et al. 2020), so the value should still be regarded as a lower limit.

[(j)]Derived from greatly depleted $NH_3$ in the upper troposphere, hence most likely not representative of N/H in the deep well-mixed atmosphere (see Atreya et al. 2020).

[(k)]Galileo probe entry site (Wong et al. (2004).

[(l)]Juno microwave radiometer - combined result of Li et al. (2024) and Li et al. (2020) using two different approaches to derive the water abundance in Jupiter's equatorial zone (values converted to protosolar here).

[(m)]Irwin et al. (2018).

[(n)]Irwin et al. (2019), in both Uranus and Neptune, the listed S/H value should be regarded as a *lower limit*, since it is derived from $H_2S$ gas detected in the 1.2–3 bar region above the $H_2S$ cloud, hence not necessarily representative of the deep well-mixed atmosphere value of S/H (Atreya et al. 2020). For somewhat deeper in the Uranus atmosphere, see footnote (i).

[(o)]Cavalie et al. (2024).

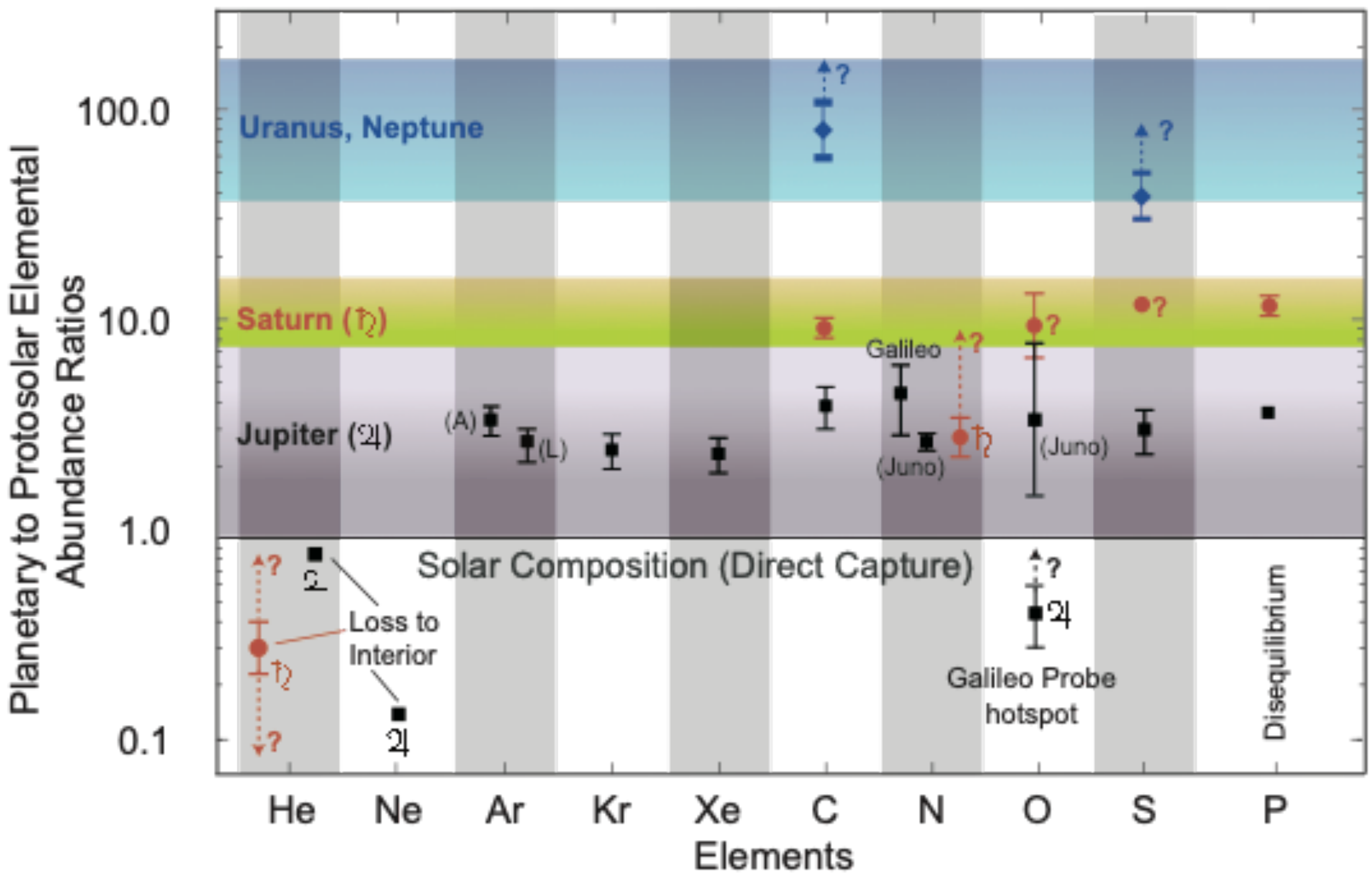

Figure 3.5 The elemental abundances in the atmosphere of Saturn, with comparison to the Sun and the other giant planets. Values marked with "?" are either uncertain (N, S, He) or may not be representative of the actual ratio in the deep atmosphere. See Footnote (c) in Table 3.3 for the possible range in Saturn's helium, shown here by upward and downward pointing arrows from the current result from CIRS. P/H is derived from $PH_3$ in the upper troposphere but could be greater in the deep atmosphere where $PH_3$ is in thermochemical equilibrium. See Table 3.3 and associated footnotes for the values, explanations, and references. [Updated from Atreya et al. 2019].

### 3.4 Volatile Delivery and Giant Planet Formation

As discussed earlier (Sec. 3.2), core accretion is the preferred model of the formation of the giant planets. The heavy elements present in their envelopes can either result from the core erosion or from the accretion of solids dragged by the protosolar nebula (PSN) gas forming the envelope. The accretion of icy solids, either in pebbles (millimetric and centimetric sizes) or planetesimals (larger than the metric size) forms, are needed in Jupiter and Saturn to account for their observed volatile enrichments. The nature and the composition of the icy phase embedded in these solids depends on the PSN thermodynamic conditions encountered during their condensation and/or drift within the disk.

If the protosolar nebula was cold enough to enable the migration of icy particles originating from the interstellar medium (ISM) toward the formation region of the giant planets without encountering any phase transition and volatile loss, then these solids should contain a mixture of volatiles embedded in amorphous ice. The key point of this scenario is that there is no fractionation during the processes of adsorption/desorption of the gaseous mixture at the surface of water ice, as shown by laboratory experiments (Bar-Nun et al. 2007). Hence, planetesimals agglomerated from amorphous grains that consecutively formed from the adsorption of a protosolar gaseous mixture should display a similar composition. This cold icy planetesimal scenario was invoked to account for the enrichment of nitrogen and argon that require temperatures <35 K for trapping, in particular, and roughly the same enrichment of all heavy elements, in general, observed in Jupiter by the Galileo probe (Owen et al. 1999), The model was subsequently generalized to all giant planets (Owen and Encrenaz 2006).

Amorphous ice experiences an amorphous-to-crystalline ice phase transition when it is heated at ~143 K (Kouchi 1990; Bar-Nun et al. 2007). At this temperature, all adsorbed volatiles are released as vapors, and water ice remains in crystalline form until it is heated at a typical temperature of ~150 K at PSN conditions. Above this value, water also becomes vapor. In the case of a warmer PSN, when ISM amorphous grains enter the disk and cross the amorphous-to-crystalline transition zone (ACTZ), i.e., the zone within which the disk temperature reaches or exceeds ~143 K, water ice crystalizes and the adsorbed volatiles are released to the disk gas phase (Mousis et al. 2019). Because the PSN slowly cools down with time, the released volatiles will form crystalline ices whose condensation temperatures are lower than the one needed to crystallize amorphous ice. These crystalline ices can consist of pure condensates forming in a 20-150 K temperature range in the PSN, depending on their abundances relative to $H_2$, and the equilibrium pressures of the considered species. Alternatively, crystalline ices can exist in the form of clathrates, which are crystalline water-based solids physically resembling ice, in which small non-polar molecules or polar molecules with hydrophobic moieties are trapped inside "cages" of hydrogen-bonded frozen water molecules (Sloan & Koh 2008). The ratio of trapped volatiles to water in clathrates is up to about 1:6 total trapped gas (Sloan & Koh 2008). Clathrates typically crystallize in the PSN at higher temperatures than those of the pure condensate forms of their encaged molecules. Their presence in the disk essentially depends on the availability of crystalline water ice in the PSN, whose abundance can significantly vary in the vicinity of its condensation front (Mousis et al. 2021), and also on the kinetics of trapping, which is not well constrained so far (Ghosh et al. 2019a, 2019b, Choukroun et al. 2019). While these two types of ices can account for the metallicities of both Jupiter and Saturn, the clathrate trapping scenario has been generally favored over the years (Gautier et al. 2001; Hersant et al. 2004; Mousis et al. 2009) because there is no confirmed evidence for the existence of primordial amorphous ice in the solar system. This preference is supported by the ESA Rosetta mission, which has shown that the composition of comet 67P/Churyumov-Gerasimenko could be more easily explained by its formation from clathrates instead of amorphous ice (Luspay-Kuti et al. 2016; Mousis et al. 2016, 2018). While the clathrate model is appealing, its validity remains to be tested. Clathrates require large amount of water to trap guest molecules of atmospheric volatiles. For example, Hersant et al. (2004) calculate a minimum of 10.5× solar water at Jupiter using Anders and Grevesse (1989) solar O/H ratio ($8.55\times10^{-4}$), or 18.4× solar $H_2O$ with the current solar O/H ratio ($4.89\times10^{-4}$) from Asplund et al. (2009), and the corresponding O/H ratio for Saturn is a minimum of 10.5× solar O/H (with Asplund et al. solar value). The equatorial value of water in Jupiter from Juno is 2.7× solar (Sec. 3.3.2, Table 3.3) and at most 7× solar by 2σ confidence interval (Li et al. 2020), which is more than a factor of two below the predictions of the above clathrate model. After the global water abundance of Jupiter is determined, it would be important to revisit the clathrate and other models of volatile enrichment including also the possibility of formation of Jupiter in a water-poor region (with C/O>1) beyond the snow line of the primordial solar nebula.

Alternative models are those advocating that the giant planets envelopes were fed from protosolar nebula (PSN) gas already enriched in heavy elements (Guillot & Hueso 2006). So far, these models are compatible with the two scenarios of disk instability and core accretion. For instance, Monga & Desch (2015) estimated that the far-ultraviolet (FUV) flux generating photoevaporation in the outer PSN enables an efficient trapping of the volatiles in amorphous water. They then proposed that amorphous grains formed in such conditions could drift toward the formation region of Jupiter. When approaching this region, grains would have crystallized and released the adsorbed volatiles, due to the increasing disk temperature. This would have progressively enriched the PSN metallicity, until it reaches Jupiter's value. The authors then suggested that Jupiter would have acquired the observed volatiles enrichments essentially from PSN gas during its growth. This idea was subsequently investigated in detail by Mousis et al. (2019), who calculated the radial profiles of volatiles enrichments around the ACTZ in the PSN, considering the inward drift of icy grains and the diffusive feedback of vapors, as a function of time (see Figure 3.6). They found that a significant enhancement of the gaseous abundances of the released volatiles was possible in a region extending over a few AU around the ACTZ, compared to their initial values.

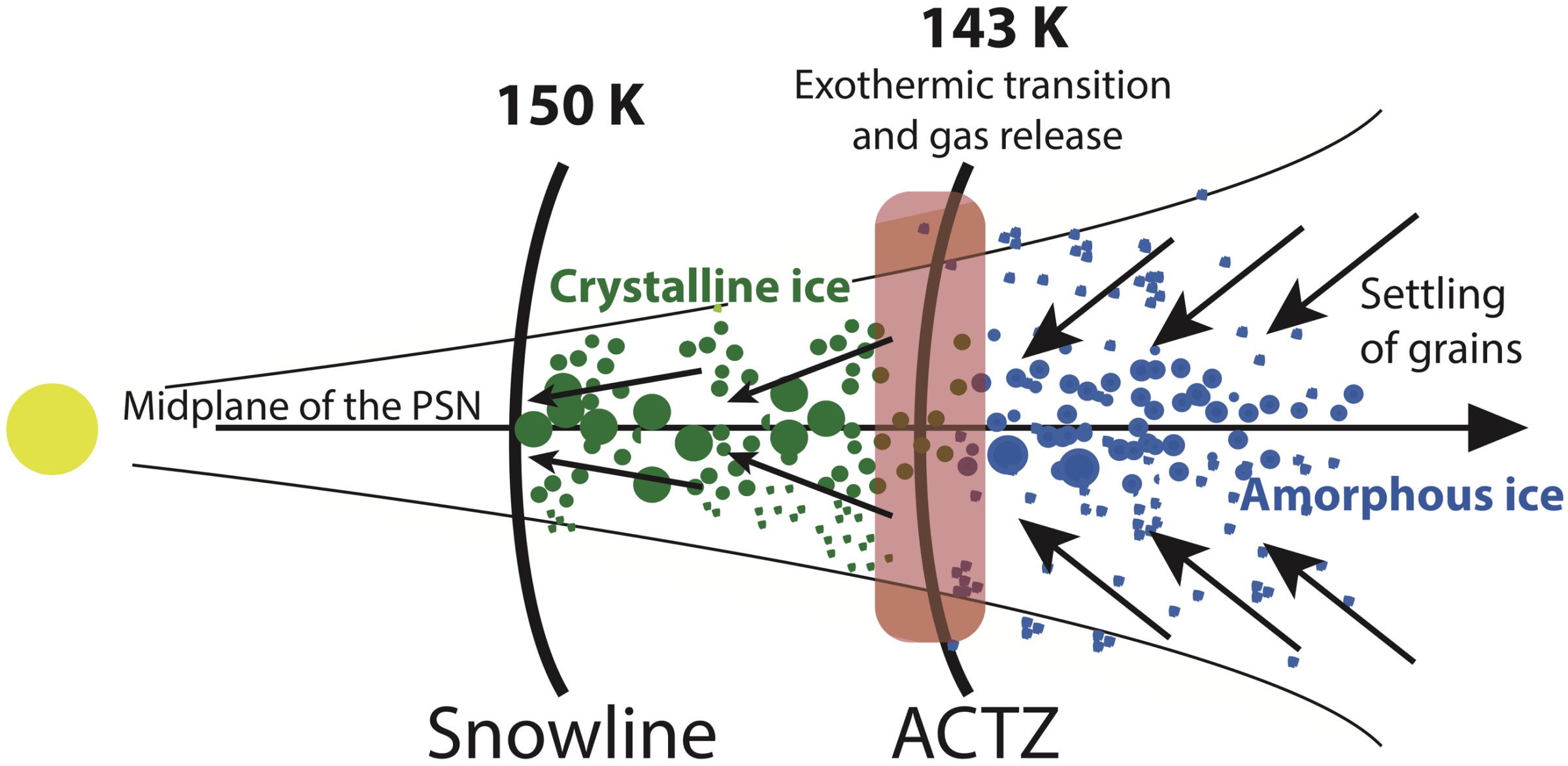

Figure 3.6 Sketch illustrating the scenario to explain a homogeneous enrichment in volatiles in the envelope of Jupiter from PSN gas only (Mousis et al. 2019). Black arrows represent the dynamical evolution of grains (sedimentation, coagulation, and inward drift). The dividing line between the amorphous and crystalline region is the ACTZ, that between ice and vapor is the snowline. Amorphous ice is stable beyond the ACTZ, crystalline ice is stable between the snowline and the ACTZ, and water remains as vapor in regions interior to the snowline. Icy particles propagate inward of the snowline and the ACTZ because inward migration is faster than sublimation for certain particle sizes.

The proposed scenario remains consistent with the water abundance measured by Juno in Jupiter's equatorial zone, which is estimated to be enriched by a factor in the 1-5 times protosolar oxygen (Li et al. 2020). However, the global abundance of water is yet unknown. The amount of water accreted in solid form by the growing Jupiter was probably limited because the planet reached and exceeded the so-called pebble isolation mass, namely the mass generating a pressure bump that ejects the drifting pebbles outside its orbit (Bitsch et al. 2018). On the other hand, Juno's high-end estimate could be consistent with Jupiter's formation around the snowline where the gaseous water abundance is found super-solar, due to the redistributive diffusion of water vapor around its vaporization location. Juno's lower-end estimate could instead be consistent with Jupiter's formation in the vicinity of the ACTZ, because of the limited amount of extra water supplied by the outward diffusion of vapor (Mousis et al. 2019). The model proposed by Mousis et al. (2019) potentially applies to Saturn as well. The only firm volatile abundance determination in its atmosphere is the carbon measurement, found close to 10 times the protosolar value (Flasar et al. 2005, Fletcher et al. 2009). Fig. 4 of Mousis et al. (2019) shows that such a high metallicity is found in the PSN at a later epoch (2 Myr vs. 0.5 Myr) than the metallicity matching the value observed at Jupiter.

### 3.5 Moons and Rings

How and when the rings and the moons formed is essential for understanding the formation of the Saturn system as a whole but also for exoplanets, for which the presence of rings is becoming an important question if not yet a reality (e.g., Kenworthy and Mamajek 2015). While some relevant data existed previously, unparalleled observations during Cassini's Grand Finale phase are allowing to address these questions in some detail.

#### *3.5.1 A Controversy About the Age of the Rings*

One of the most remarkable achievements of Cassini's Grand Finale is the *measure* of the mass of the rings. By radio-tracking the spacecraft during 6 of its free-falls between the planet and the rings, Iess et al. (2019) were able to disentangle the gravity accelerations from the planet and from the rings, as they pull in different directions. While the detailed analysis of the gravity field of Saturn provides crucial information on its internal structure, the mass of the rings is a precious constraint to all formation and evolution models. Iess et al. (2019) find a ring mass of $(1.54 \pm 0.49) \times 10^{19}$ kg, or ~0.4 times the mass of Mimas, the smallest of the spherically shaped moons of Saturn, with 400 km diameter.

The first published interpretation of this number is the age of the rings. Indeed, we know from Zhang et al. (2017a,b) that the rings are made mostly of water ice, with only <~1% of other materials (organics, metals, rocks… hereafter called "dirt"). The total mass of the rings therefore gives the total mass of dirt that is present in the rings. But the rings are constantly bombarded by dirty micrometeorites, interplanetary dust grains, etc. This flux has also been measured by the Cassini mission and agrees with older estimates so that all this dirt can be brought to the rings in less than 100 million years. The link between the measured mass and the age relies on the hypotheses: (i) that the ring mass is constant with time (ii) that the bombardment rate is constant with time and (iii) that the rings retain all the pollution they receive. While these three hypotheses are very plausible, the formation of such a ring system only a hundred million years ago is not. Ćuk et al. (2016) argue that a dynamical instability among Saturn's satellites must have occurred in the recent past, possibly leading to catastrophic collisions among satellites, and that the debris could have generated the present system. But the details of this generation remain to date elusive, and the age of some craters on Rhea and Dione, estimated to be more than 500 Myrs (Dalla-Ore et al. 2015, López-Oquendo et al. 2019, Kirchoff et al. 2018) seems to contradict this scenario. Dubinski (2019) proposes that the rings formed from the collision of a 20 km comet with a satellite twice the size of Mimas, but although this collision could happen at any time in the history of the solar system, it is much less likely to be recent. On the other hand, the present mass corresponds to the one obtained after a few Gyrs of viscous evolution of the rings, independent of the initial mass (Salmon et al. 2010, Crida et al. 2019), which seems to support old rings, maybe as old as Saturn. In this case, one of the above three hypotheses must be incorrect. Based on the detection during the Grand Finale of nanograins of silicates flowing from the rings towards Saturn (Hsu et al. 2018), together with the discovery of large amounts of hydrocarbons in the upper atmosphere of Saturn (Waite et al. 2018), Crida et al. (2019) argue that the rings may be self-cleaning and eliminate pollution as they receive it. However, no such cleaning mechanism has been proposed so far, and the physics of hypervelocity impacts of micrometeorites on porous ice blocs is still vastly unknown. In addition, the effects of the bombardment on the dynamics of the rings, and not only on their composition, may play an important role in their evolution, so that the question of the age (and therefore of the origin) of the rings remains open.

### *3.5.2 A Fast Migration of Titan*

A controversial result has been published recently, which shed a new light on the satellite system as a whole. Using precise radio tracking of the Cassini spacecraft during its encounters with Titan, combined with astrometry of all the satellites, from the Cassini mission, Lainey et al. (2020) found that Titan is migrating away from Saturn by 11 cm per year. Outwards migration of a satellite is a natural consequence of tidal dissipation in the planet. But in the traditional view, the efficiency of this phenomenon drops sharply with distance, so that Titan's shift should be hardly detectable. In contrast, this result—if correct—shows that tidal dissipation inside Saturn is extremely efficient at Titan's frequency. Because no uniform tides model would predict such a high rate, this suggests that tidal dissipation does depend on the excitation frequency. Moreover, the data suggest that the migration time (defined as $r/v_{mig}$ where r is the distance to Saturn and $v_{mig}$ the migration speed) is the same for all the satellites (Figure 3.7). This is consistent with a recent model of tides called resonance locking, in which the tides from satellites excite oscillation modes inside Saturn (Fuller et al. 2016). If the excitation frequency is close to the proper frequency of the mode, its amplitude rises, as well its dissipation. This promotes a fast outwards migration of the satellite, whose orbital frequency then decreases. Now, if the proper frequency of the mode decreases also, because of the internal evolution of Saturn, it catches up with the satellite. The migration speed is thus eventually set by Saturn's evolution time. This time is the same for all the satellites, and always of the order of the age of Saturn. This implies that all the satellites have the same migration time, and that they all originate from r=0 at t=0.

The above description of the resonance locking model is simplistic, but it shows that all the dynamics of the system must be re-examined. In particular, in this framework, the satellites' orbital period ratios are constant. There are no more resonance crossing (as assumed by Ćuk et al. 2016) nor orbit crossing and collisions (as assumed by Charnoz et al. 2011 and Crida & Charnoz 2012). Even Titan, like the other moons interior to its orbit, could be originating from the rings spreading beyond the Roche radius instead of having formed in the circumsaturnian disk as classically assumed.

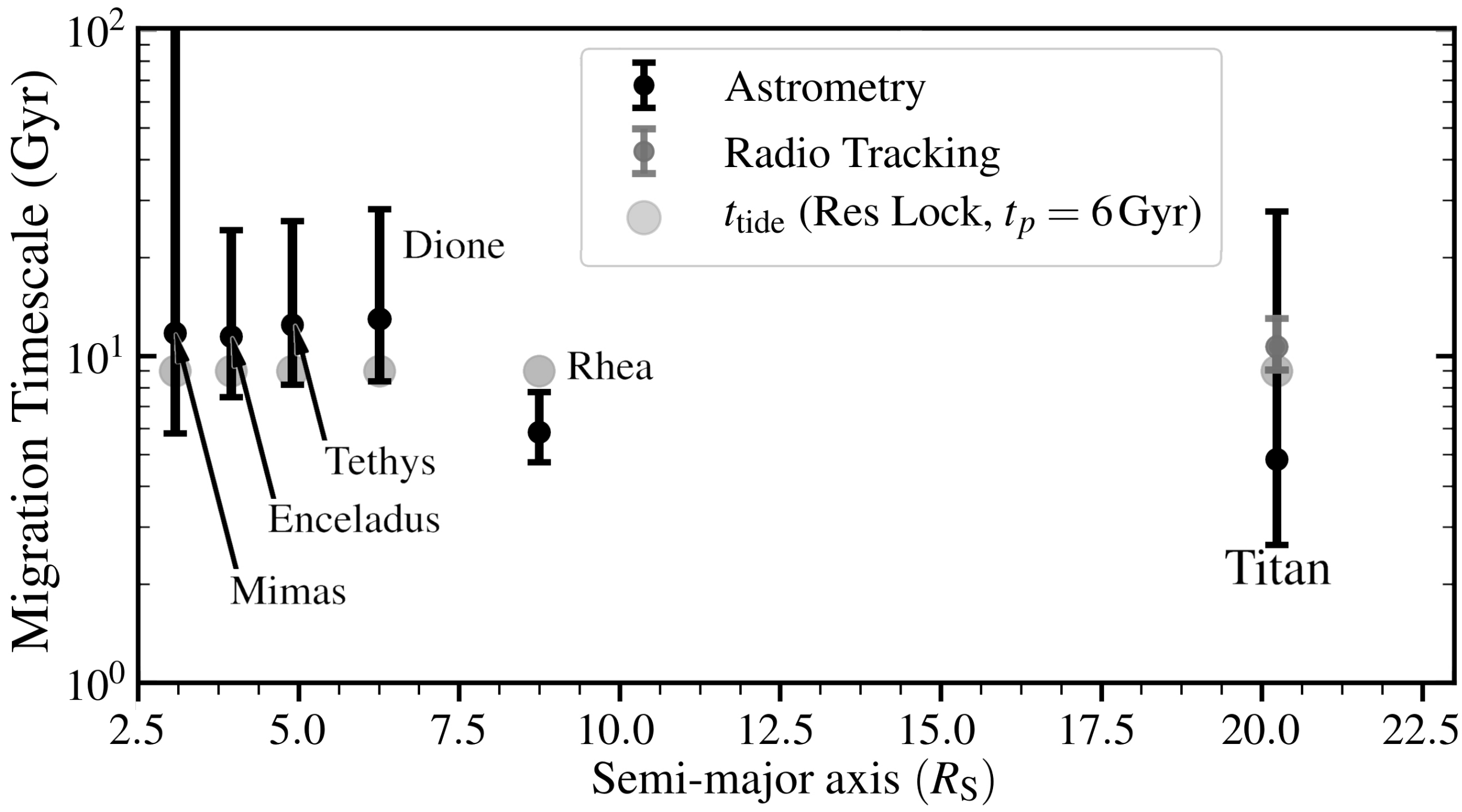

Figure 3.7 Measured migration time ($r/v_{mig}$) of the mid-sized moons, thanks to astrometry (black) and radio-tracking of Cassini (dark grey, for Titan). [Adapted from Lainey et al. 2020, Fig. 2].

### *3.5.3 Who Came First? The Rings or the Moons?*

In the end, we are left with a chicken-and-egg dilemma. Between the rings and the satellites, which gave birth to which?

Dynamics suggest that the rings are old and gave birth to all the regular satellites of Saturn. Composition arguments suggest that the rings are young and thus are probably coming from the recent destruction of satellite(s). In any case, the link is strong, and has been reinforced by Cassini VIMS measurements that the D/H ratio in the rings is consistent with that of Enceladus' surface (Clark et al. 2019). This can be understood either way: Enceladus may be coming from the rings, or its plumes may be coating the rings with fresh ice (making them look bright and younger than they actually are), or they could both come from the same catastrophic collision in the system.

In the case where the rings are primordial, and generated all the regular satellites, the most convincing formation model so far is the tidal destruction of a satellite migrating in the circum-planetary disc (CPD) inside the Roche radius (Canup 2010)! In the end, a better understanding of the formation of satellites in the CPD is needed, but this may only constrain the precursor of the rings, which are themselves the precursors of Titan, Rhea, Dione, Tethys, Enceladus, Mimas, etc. See also Sec. 3.2 for related discussion.

## 3.6 Exoplanet Perspective

### *3.6.1 Demographics of Exo-Saturns*

It is interesting to place Saturn in an exoplanet perspective. There is currently no exact Saturn analogue in exoplanets in terms of its mass, radius, temperature, and orbital separation. In particular, the 10 AU orbit of Saturn is beyond the reach of transit or Radial Velocity (RV) exoplanet surveys and the Saturn-Sun flux contrast is below the sensitivity of direct imaging surveys (see e.g., Fischer et al. 2014 for review of exoplanet detection methods). One exception is microlensing, which enabled the detection of a Jupiter-Saturn analog system with slightly smaller masses and smaller separations (Gaudi et al. 2008). However, numerous exoplanets with similar masses and sizes as Saturn are known in close-in orbits, i.e., highly irradiated "hot Saturns", discovered by transit and/or RV surveys. It is also interesting to ask: what defines an exo-Saturn? Unlike the solar system, the exoplanet population does not have a clear gap between ice giants and gas giants, with a unique place for Saturn. Instead, exoplanets occupy a continuum in masses and radii over a vast range that is not encountered in the solar system: from Earth-size planets to super-Jupiters. Within this range, there are also many planets with masses similar to Saturn but with diverse radii and others with Saturn-like sizes but very different masses. Such diversity is accorded by a combination of different possible formation, migration and evolutionary processes and equally diverse atmospheric processes resulting from the high irradiation in close-in orbits. Interestingly though, Hatzes & Rauer (2015) find that in a mass−density scatter plot of known exoplanets, a sharp transition occurs exactly at Saturn's mass, where the density starts to increase almost linearly with the mass up to 60 Jupiter masses, before decreasing again. Hence, these authors argue that Saturn's mass should be the definition of the lower mass for a "*giant gaseous planet*", making Saturn a particularly interesting case study for exoplanets. Here, we nominally refer to "exo-Saturns" or "hot Saturns" as planets with masses like that of Saturn, including those between Neptune and Saturn masses.

### *3.6.2 Atmospheric Chemical Compositions of exo-Saturns*

In recent years, transit spectroscopy has led to chemical detections in several exo-Saturns. The detections are primarily of $H_2O$ and the alkali metals Na and/or K (e.g., Chen et al. 2018, Nikolov et al. 2018, Wakeford et al. 2018, Kirk et al. 2019, Colon et al. 2020), as shown in Figure 3.8. The $H_2O$ detections have been carried out in the 1.1-1.7 μm range using the near-infrared spectroscopy capability of the Hubble Space Telescope (HST) WFC3 spectrograph. The Na/K detections at optical wavelengths are made with large ground-based telescopes (e.g. Nikolov et al. 2018 and Chen et al. 2018) or with the HST STIS spectrograph (e.g. Sing et al. 2016). In some exo-Saturns atomic He has also been discovered using high-resolution transmission spectroscopy of the He I triplet line at 10833 Å (e.g., Spake et al. 2018, Nortmann et al. 2018).

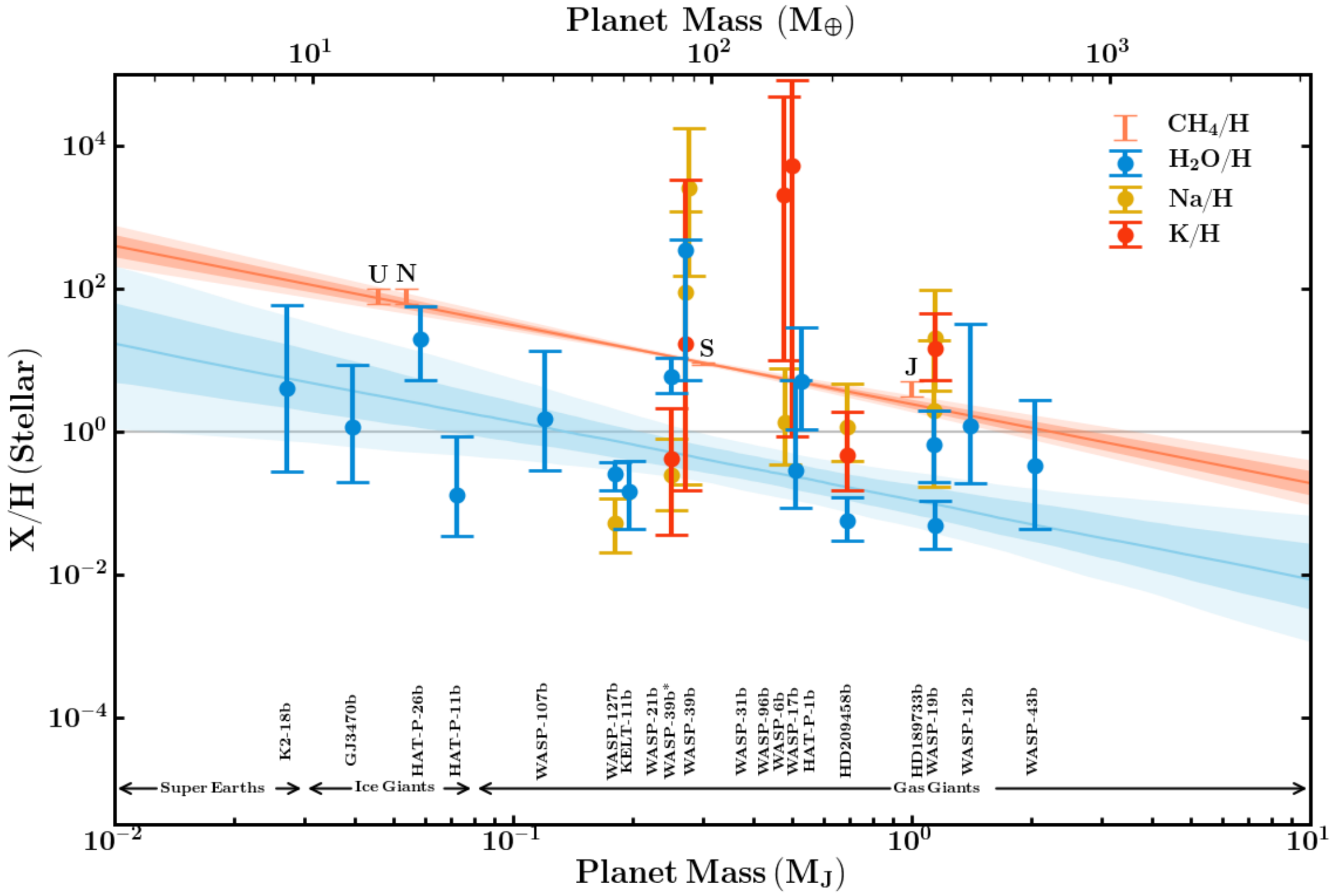

Figure 3.8 Atmospheric abundances of transiting exoplanets vs. planet mass. The $CH_4$ abundances of the solar system giant planets are shown in coral orange, and the corresponding trend is also shown along with the 1- and 2- sigma confidence regions. The exoplanet abundances are shown for $H_2O$ (blue), Na (orange) and K (red). J: Jupiter, S: Saturn, U: Uranus, N: Neptune. [Adapted from Welbanks et al. 2019]. Alkali metal abundances of of a hot Saturn and a hot Jupiter with cloud and haze free atmospheres, not shown in this figure, range from sub-solar to super-solar (see text).

Neither $H_2O$ nor the alkali metals (Na, K) have been detected on Saturn, where all these species condense out because of the planet's low atmospheric temperature. On the other hand, while $CH_4$ and $NH_3$ gases are strongly detected in Saturn's low temperature environment, they are not detected in hot exo-Saturns. Thus, the chemical signatures of exo-Saturns provide complementary information to what is presently known on Saturn. Similarly, the atmospheres of exo-Saturns show a range in cloud cover, from cloudy to cloud-free (e.g., Sing et al. 2016, Nikolov et al. 2018, Pinhas et al. 2019, Welbanks et al. 2019). However, even for the planets that display cloudy atmospheres, the cloud composition is expected to be made of high-temperature condensates, e.g., silicates and other metal-rich species (e.g., Sudarsky et al. 2003, Morley et al. 2013), rather than the low-temperature $H_2O$, $NH_3$ and $NH_4SH$ clouds expected on Saturn and the other solar system giants (see Fig. 2.9, Atreya et al. 2019). The nature of clouds and hazes in the hot upper atmospheres of exo-Saturns can provide important clues about the thermochemical processes in the deep atmosphere of Saturn, where such condensates of rock and metal are predicted to occur at high temperatures and pressures as shown in Figure 3.9. Juno's microwave radiometer (MWR) measurements have successfully probed the deep atmosphere of Jupiter, where free electrons upon the ionization of alkali metals provide the excess opacity required to explain the data. Preliminary analysis of the MWR data in the equatorial zone indicated that the alkali metals are greatly depleted, by a factor of $10^2$-$10^5$ relative to solar at ~1 kilobar level (Bhattacharya et al. 2023). Further modeling finds a less severe depletion of 0.1× solar in Na and K, since the free electrons attach to atmospheric gases, resulting in the formation of anions, particularly $HS^-$ and $Cl^-$, whose abundance is calculated to be ten times greater than free electrons (Aglyamov et al. 2024). The sub-solar abundance of the alkalis is in contrast with other heavy elements in Jupiter all of which have a super-solar abundance (Sec. 3.3.4). Considering that the MWR can sense down to only about one kilobar level, the depletion of alkali metals at that level does not necessarily imply a sub-solar elemental abundance of Na and K in Jupiter as a whole because of the possibility of their removal in the interior by thermochemistry or sequestration at 10s of kilobar, processes that are poorly understood in the absence of relevant laboratory data. Until then, 0.1× solar should be regarded as a lower limit on the elemental abundance of the alkali metals in Jupiter. A comparison of the alkali metal abundance in the other gas giant, Saturn, would be highly illuminating. Microwave remote sensing from a future spacecraft mission at Saturn will have an excellent chance of determining their abundance in Saturn, especially since the synchrotron radiation effects at Saturn are minimal compared to Jupiter. Alkali metal abundance can be an important addition to Saturn's suite of other heavy elements determined by a future entry probe (Sec.3.3.4). Hot Jupiters/Saturns can also provide useful insights into the alkalis as discussed below.

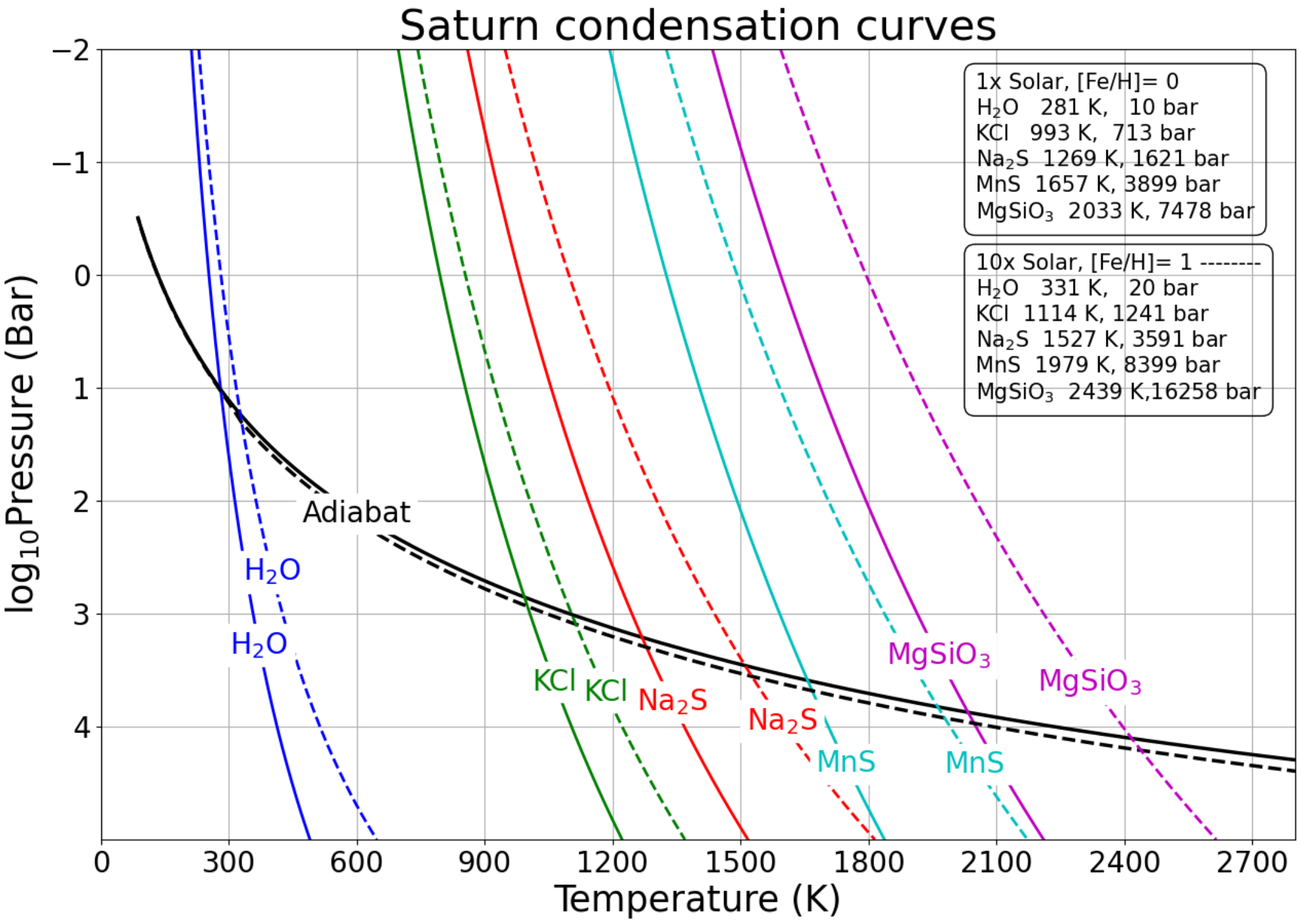

Figure 3.9. Condensation curves of cloud-forming species for a Saturn-like atmosphere. At low temperatures in the observable atmosphere, e.g., for Saturn, the clouds are composed primarily of $H_2O$, along with lower-temperature condensates at atmospheric pressures <25 bars (Atreya et al., 2019, Fig. 2.9). For “hot Saturn” exoplanets on the other hand, the clouds can be composed of various high-temperature condensates of alkali metals ($Na_2S$, KCl) and rock/refractory material ($MgSiO_3$, MnS, etc.) as shown. The adiabat of Saturn is shown in black for reference (metallicity, Fe/H, in the inset is defined as $Fe/H = \log_{10}[(N_{Fe}/N_H)_{star}/(N_{Fe}/N_H)_{sun}]$, so that Fe/H=0 means 1× solar (solid black line) and Fe/H=1 means 10× solar (dashed black line), $N_{Fe}$ and $N_H$ are number densities of iron and hydrogen, respectively). At highly depleted abundances above clouds would form at lower temperatures (at 0.01× solar abundances, for example, $H_2O$ condenses at approximately 222 K, 5 bar; KCl at 830 K, 380 bar; $Na_2S$ at 968 K, 640 bar; MnS at 1270 K, 1584 bar; and $MgSiO_3$ at 1550 K, 3056 bar. Condensation equations are from Lodders, 2003; Visscher et al, 2006, 2010; Morley et al., 2012), and solar abundances from Asplund et al. (2009).

The chemical abundance measurements in exo-Saturns may provide complementary constraints on the atmospheric metallicity to those in Saturn. The metallicity in Saturn's atmosphere, derived from the $CH_4$ abundance, is estimated to be C/H ~ 10×solar, as discussed in previous sections. While the C/H ratios are not known reliably for exoplanets, the $H_2O$ abundances in hot Jupiters and hot Saturns are found to be nearly solar or sub-solar (Madhusudhan et al. 2014, Kreidberg et al. 2014, Barstow et al. 2017, Pinhas et al. 2019, Welbanks et al. 2019, Colon et al. 2020). As shown in Fig. 3.8, the $H_2O$ abundances are generally lower than those predicted based on the solar system mass-metallicity trend (Pinhas et al. 2019, Welbanks et al. 2019). On the other hand, the Na/K abundances estimated for only a handful of planets appear to be consistent with significantly super-solar metallicities (Chen et al. 2018, Welbanks et al. 2019). However, the uncertainties are large (Figure 3.8), so that sub-solar abundances cannot be ruled out. Recent observations of giant exoplanets with clear or nearly clear atmospheres, which do not suffer from ambiguities resulting from modeling of radiative transfer through cloudy/hazy atmospheres, add to the confusion: while the Na/K abundance on hot Jupiter HAT-P-1b is found to be sub-solar (Chen et al. 2022), it is solar to greatly super-solar on hot Saturn WASP-96b (Nikolov et al. 2022).

Abundance estimates of other elements are also being debated in the literature due to differing datasets and abundance analyses (e.g., Wakeford et al. 2018, Kirk et al. 2019, Welbanks et al. 2019). These trends suggest the possibility that the $H_2O$ abundances measured in hot Jupiters and hot Saturns may not be representative of the true atmospheric metallicity and that the oxygen abundance may be relatively less abundant compared to the other elements, such as Na/K or C. A super-solar C/O ratio (e.g. C/O ~ 1) can cause a low $H_2O$ abundance at high temperatures (≥1200 K) even if the overall O/H and C/H ratios may be high; most of the oxygen would be bound in CO under these conditions (Madhusudhan 2012, Moses et al. 2013). An important caveat about the above trends and the sub-solar to super-solar range of Na/K is that they are based on a relatively small number of planets and their measurement uncertainties are relatively large. Only future observations, e.g., with JWST or large ground-based high-resolution spectrographs, would allow more stringent constraints on the C/H and C/O ratios and the Na/K abundances in hot Saturns and assess their similarities with Saturn.

### *3.6.3 Implications for Planetary Formation*

The current constraints on the chemical abundances of exo-Saturns allow multiple formation and migrations scenarios for their origins. In the solar system, the super-solar atmospheric compositions of Saturn, and Jupiter, have traditionally been explained by their formation via core accretion (Pollack et al. 1996, Owen et al. 1999, Atreya et al. 2019); see Sec. 3.2. A caveat to note is that the global abundance of $H_2O$ is presently unknown in Saturn. However, the observations of generally low $H_2O$ abundances in exo-Saturns and hot Jupiters in general have motivated considerations of diverse formation and migration pathways. As alluded to above, the generally low $H_2O$ abundances could be caused either by low absolute oxygen abundances, i.e., low O/H, or by high C/O ratios.

Several studies have explained the possibility of high C/O ratios in gas giants formed outside the CO snow line and migrating inward without further accretion of solids in the disk, e.g., by disk-free migration mechanisms (Oberg et al. 2011, Madhusudhan et al. 2014). High C/O ratios can also be caused due to formation by pebble accretion where the solids are locked in the core and the gas occupying the envelope has a high C/O ratio when formed outside the CO snow line (Madhusudhan et al. 2017). In both the above scenarios, the high C/O in the atmosphere is accompanied by a low (sub-solar) metallicity. On the other hand, a combination of high C/O ratios as well as a high metallicity is possible when the gas in the disk is metal-enriched near the snowlines due to pebble drift (Booth et al. 2017). Regardless of the specific formation scenario, if the atmosphere has a high C/O ratio it is a strong indicator that the planet acquired the majority of its gaseous envelope in the outer regions of the disk, close to or beyond the CO snowline and migrating to its present close-in orbit once the disk has dissipated, i.e., without significant solid accretion during migration. Such a scenario may not be entirely consistent with the origin of Saturn in the solar system whose present location in what would have been the outer disk implies that its composition could have inherited entirely in situ (see also Sec. 3.2 and 3.4). While the interpretation of current volatile abundances of exo-Saturns leaves room for different formation scenarios, the nominal trend of atmospheric metallicity increasing with decreasing planet mass generally favors the core accretion model. The occurrence frequency of giant exoplanets also increases with increasing stellar metallicity, which is a strong evidence in support of the core accretion model since the higher heavy element or metal content facilitates core formation (Johnson et al. 2010; Howard et al. 2013; Mortier et al. 2013).

### 3.7 Summary and Future Directions

We have discussed the various scenarios of the formation of giant planets to gain an insight into the origin and evolution of Saturn. Available constraints on models, including the substantially greater enrichment of heavy elements observed at Saturn compared to Jupiter coupled with comparable core mass, argue strongly in favor of the core accretion model, as opposed to the disk instability model. However, data on the abundances of core-forming heavy elements in Saturn are quite sparse compared to Jupiter. A highly precise determination of the C/H ratio has been made from $CH_4$, but the same cannot be said about N/H or S/H, and several other key elements are yet to be measured. $NH_3$ has been measured only in the equatorial region down to ~3 bar level, yielding an N/H ratio that is a factor 3 smaller than the C/H ratio. The difference is so large, that one must wonder whether the observed depletion in $NH_3$ is representative of the deep well-mixed atmosphere globally or is it the result of some complex convective-thermodynamic process, such as the mushball formation model to explain similar depletion of $NH_3$ at Jupiter observed by Juno. If the relatively low N/H value of Saturn were indeed global, it would have significant implications for the formation and evolution models of the gas giants. The helium elemental abundance poses a similar challenge. While Saturn's excess luminosity argues for a low He/H ratio, present values range from very low to very high. That has profound implications for the interior processes, heat balance and Saturn's evolution. Exo-Saturns can provide some insight into possible elemental abundances at Saturn, but those that can be studied are in close proximity to their parent stars, and such "hot Saturns" may not be true analogs of Saturn in the solar system. Cassini-Huygens has dramatically enhanced our understanding of the entire Saturn system including the planet's atmosphere, rings and the moons but has also posed questions that remain unanswered. Juno has provided an unprecedented look into the workings of one gas giant and only a future mission to Saturn can reveal whether those ideas apply also to the other gas giant in our solar system. For the origin and evolution of Saturn, abundances and the isotopic ratios of the noble gases, He, Ne, Ar, Kr and Xe, and the elements sequestered in condensible gases - N, S and O - are key. Helium is also key to understanding Saturn's luminosity and the interior. Only entry probes are capable of measuring the noble gases, using the proven technique of mass spectrometry (Niemann et al. 1998; Mahaffy et al. 1998, 2000). Their measurements can be made at relatively shallow depths of a few to ten bars and the abundances even at a single location of the probe entry site are expected to be representative of their global values since noble gases are chemically inert and they do not condense at Saturn. Entry probes can measure the condensible volatiles also, but require deployment to very deep levels, which at Saturn could be several tens of bars or more, in order to determine O/H from well-mixed water. Another difficulty with measuring well-mixed condensible gases from probes is the evidence from Juno for strong spatial variations in the ammonia abundance even below the modeled cloud base. Current probe technology prevents the determination of elemental abundances from measurements of condensible gases at Saturn, especially O/H and N/H, unless their well-mixed abundances are substantially sub-solar, contrary to expectations. On the other hand, precise measurement of some of their isotopic ratios would be feasible at relatively shallow depths, especially using the tunable laser spectrometry technique (Webster et al. 2024). Isotopic ratios are important for understanding the volatile origin and delivery. Microwave radiometry from an orbiter for global mapping of $NH_3$ and $H_2O$ to great depths, as on Juno, would be an ideal complement to the probe. The possibility of deriving O/H from CO appears promising also, so should be investigated more thoroughly along with related disequilibrium species. To conclude, in situ measurements with an entry probe combined with critical orbital observations on a future Saturn missions and their comparison with corresponding existing data of Jupiter is essential for developing robust, comprehensive and unambiguous models of the formation and evolution of Saturn, the giant planets, and the solar system, and by extension, the extrasolar systems.

## Acknowledgments

SKA thanks Tarun Kumar and Pranika Gupta for help with the illustrations, Abigail Stautberg for bibliography and many colleagues on the Cassini, Juno and the Galileo teams for their insights during the course of preparing the manuscript. We are especially grateful to Paul Mahaffy for illuminating discussions on the xenon isotope ratios measured by the Galileo Probe at Jupiter. SKA was supported in part by the University of Michigan's Discretionary Fund and by the Juno Mission through subcontract 699056KC from the Southwest Research Institute; JIL was supported by Juno subcontract D99069MO.